\documentclass[a4paper, 12pt]{article}
\usepackage{a4}
\usepackage{amsfonts}
\usepackage{amssymb}
\usepackage{cite}
\usepackage{epsfig}
\usepackage[all]{xy}
\usepackage{amsmath}
\usepackage{cite}

\DeclareMathOperator{\ch}{ch}

\DeclareMathOperator{\rank}{rk}

\author{}

\newcommand{\cL}{{\mathcal{L}}}
\newcommand{\be}{\begin{equation}}
\newcommand{\ee}{\end{equation}}
\newcommand{\ba}{\begin{array}}
\newcommand{\ea}{\end{array}}
\newcommand{\bea}{\begin{eqnarray}}
\newcommand{\eea}{\end{eqnarray}}

\newcommand{\3}{{\bf 3}}
\newcommand{\4}{{\bf 4}}
\newcommand{\5}{{\bf 5}}
\newcommand{\6}{{\bf 6}}

\newcommand{\ov}{\overline}
\def\IR{\relax{\rm I\kern-.18em R}}

\def\IP{\relax{\rm I\kern-.18em P}}
\def\inbar{\vrule height1.5ex width.4pt depth0pt}
\def\IC{\relax\,\hbox{$\inbar\kern-.3em{\rm C}$}}

\def\K3{{\bf K3}}

\def\a{\alpha}
\def\b{\beta}

\def\ov{\overline}

\def\n2d{\cN_{V^*}^{\otimes 2}}

\def\IC{\mathbb{C}}

\def\IR{\mathbb{R}}

\def\IP{\mathbb{P}}

\def\cN{{\mathcal N}}

\def\cO{{\mathcal O}}
\def\cC{{\mathcal C}}

\def\cL{{\mathcal L}}
\def\nn{\nonumber}

\def\ch{\mbox{ch}}

\def\to{\rightarrow}

\begin{document}

\title{
\begin{flushright} \vspace{-2cm}
{\small MPP-2006-179 \\
 \small RUNHETC-2006-33\\
 \small UPR-1168-T\\
\vspace{-0.35cm}
hep-th/0612039} \end{flushright}
\vspace{2.5cm}
Massless Spectra of Three Generation \\
U(N) Heterotic String Vacua   
\quad
}
\vspace{1.0cm}
\author{\small Ralph~Blumenhagen$^{1}$, Sebastian Moster$^{1}$, Ren\'e Reinbacher$^{2}$
and  Timo Weigand$^{3}$}

\date{}

\maketitle

\begin{center}
\emph{$^{1         }$ Max-Planck-Institut f\"ur Physik, F\"ohringer Ring 6, \\
  80805 M\"unchen, Germany } \\
\vspace{0.1cm}
\emph{$^{2        }$ Department of Physics and Astronomy, Rutgers University, \\
     Piscataway, NJ 08855-0849, USA } \\
\vspace{0.1cm}
\emph{$^{3        }$ Department of Physics and Astronomy, University of Pennsylvania, \\
     Philadelphia, PA 19104-6396, USA } 
\vspace{0.2cm}

\tt{blumenha,moster @mppmu.mpg.de, rreinb@physics.rutgers.edu, timo@sas.upenn.edu}
\vspace{1.0cm}
\end{center}
\vspace{0.7cm}

\begin{abstract}
\noindent  
We provide the methods to compute the complete massless spectra of a 
class of recently
introduced supersymmetric $E_8\times E_8$ heterotic string models which invoke  vector bundles
with $U(N)$ structure group on simply connected Calabi-Yau manifolds and which yield flipped SU(5) and MSSM string vacua of potential phenomenological interest.
We apply Leray spectral sequences in order to derive
the localisation of the cohomology groups
$H^i(X,V_a\otimes V_b)$, $H^i(X,\bigwedge^2 V)$ and $H^i(X,{\bf S}^2 V)$ for vector
bundles defined via Fourier-Mukai transforms on elliptically fibered Calabi-Yau manifolds. 
By the method of bundle extensions we define  a stable $U(4)$ vector
bundle leading to the first flipped $SU(5)$
model with just three generations, i.e.
without any vector-like matter. Along the way, we propose the notion of $\lambda$-stability for heterotic bundles.

\end{abstract}

\thispagestyle{empty}
\clearpage

\tableofcontents

\section{Introduction}

Since its discovery in 1985 the heterotic string \cite{Gross:1984dd} has been considered as a promising
candidate to yield four-dimensional string vacua whose low-energy effective
action resemble the Standard Model (SM) of particle physics or an extension thereof.
Different constructions, based on the subsequent discovery of D-branes, such 
as intersecting D-brane models provide an alternative way to realize
many of the Standard Model properties in concrete four-dimensional
string vacua.\footnote{For references see e.g. the latest reviews \cite{Blumenhagen:2005mu} or \cite{Blumenhagen:2006ci}.} 
Whereas these latter constructions are well-suited to directly
yield the Standard Model gauge symmetry at the string scale, for GUT-like
theories the $E_8\times E_8$ heterotic string
seems to be particularly natural. 

In the last couple of years there has been a revival of attempts to construct
realistic $E_8\times E_8$ string vacua on Calabi-Yau manifolds. In fact using advanced  techniques
for the construction of vector bundles, models have been found with the MSSM massless charged particle spectrum.
Some of them are completely supersymmetric \cite{Bouchard:2005ag,Bouchard:2006dn}
whereas others include an explicit
supersymmetry breaking hidden $E_8$ or M-theoretic bulk sector \cite{Braun:2005ux,Braun:2005nv}. 

The philosophy of these constructions, pioneered in \cite{Witten:1985bz}, is to embed an $SU(4)$ or $SU(5)$ bundle into one $E_8$ which first gives
rise to an $SO(10)$ and $SU(5)$ observable GUT model, respectively. Due to the absence
of candidate GUT Higgs fields, this gauge symmetry has to be broken by discrete Wilson
lines. In fact, most of the effort has gone into the investigation  of Calabi-Yau manifolds
admitting  non-trivial discrete Wilson lines and the construction of appropriate
equivariant vector bundles \cite{Donagi:2000fw,Donagi:2000zs,Braun:2004xv}.

Based on the methods developed in \cite{Blumenhagen:2005ga,Blumenhagen:2005pm}, an alternative to this procedure has recently been presented in \cite{Blumenhagen:2006ux}. There it is shown that by allowing
also vector bundles with $U(N)$ structure groups, the massless spectrum
can contain GUT Higgs scalars.  This approach therefore circumvents
the necessity of working on manifolds with non-trivial fundamental group and opens up the way for heterotic model building on much more general background manifolds.
In particular, the technology for the construction of stable vector bundles
on elliptically fibered Calabi-Yau manifolds \cite{Friedman:1997yq,donagi-1997-1,Friedman:1997ih} can directly be employed.

Concretely, the approach of \cite{Blumenhagen:2006ux} provides two alternative strategies. The first option consists of embedding a vector bundle with structure group $SU(4)\times U(1)$
into the first $E_8$. This engineers the observable gauge symmetry
$SU(5)\times U(1)$. Under the $U(1)$ the prospective SM particles carry exactly
the charge known from $U(1)_X$ in the flipped $SU(5)$ GUT scenario \cite{Barr:1981qv,Derendinger:1983aj}. However, without further refinements the $U(1)$ becomes massive  due to the Green-Schwarz
mechanism. 
To remedy this one can embed in addition a line
bundle into the second $E_8$ factor yielding an observable $E_7\times U(1)$ 
gauge symmetry. Under certain conditions on the vector bundles one linear
combination of the two $U(1)$ factors from the first and the second $E_8$
remains massless, eventually giving rise to a supersymmetric flipped $SU(5)$ GUT model.
The role of the GUT Higgs pair is played by the component in the $[{\bf 10 - \overline {10}}]$ neutral under the SM gauge group. One physical motivation to study the resulting models  are the known phenomenologically attractive field-theoretic features of the  flipped $SU(5)$ scenario \cite{Antoniadis:1987dx,Antoniadis:1989zy}.\footnote{Recent alternative constructions of flipped $SU(5)$ vacua in various contexts include \cite{Chen:2005cf,Chen:2006ip,Cvetic:2006by,Kim:2006hw,Dundee:2006ii}.} These include a high degree of proton stability, among others due to a natural solution to the doublet-triplet splitting problem, and distinguish the flipped from the non-flipped $SU(5)$ models. Whereas in the purely field-theoretic flipped $SU(5)$ model the GUT scale value of the $U(1)_X$ gauge coupling is in principle a free parameter, 
in the stringy flipped $SU(5)$ of \cite{Blumenhagen:2006ux} the three tree-level  gauge couplings are uniquely determined and do
not unify at the string or GUT scale. Nonetheless gauge coupling unification can in principle be achieved by a suitable tuning of the stringy threshold corrections. Moreover, there 
appear, in general,  exotic massless states, which turn out to be all  vector-like as soon as one requires that the $U(1)_X$ stays massless. It is important to note that the presence of these exotics is by no means a definite prediction of this string theoretic realisation of flipped $SU(5)$ since they can well be avoided by a suitable choice of bundle data. In fact, it is the main result of the present paper to exemplify that this is indeed possible.

The second option studied in \cite{Blumenhagen:2006ux} is to embed an $SU(5)\times U(1)$ bundle into one $E_8$ 
and a second line bundle into the other $E_8$.  This yields string vacua with
just the Standard Model gauge symmetry and, again, only very few non-chiral exotic matter states, which may or may not be present depending on the details of the compactification data.
In \cite{Blumenhagen:2006ux}, it was also carried out a successful computer search for chiral three-generation flipped $SU(5)$ and direct MSSM models on elliptically fibered Calabi-Yau manifolds, where the base was allowed to be
either a Hirzebruch surface or a del Pezzo surface with $r\le 4$.

The final aim of this paper is to continue the model search of \cite{Blumenhagen:2006ux} and to demonstrate the existence of string vacua of the two types described above and with as little vector-like exotic matter as possible. These could then serve as the starting point for concrete phenomenological studies.  
Technically, our models will be based on the extension of spectral cover bundles with structure group $U(N)$ on elliptically fibered Calabi-Yau manifolds, as pioneered in \cite{Friedman:1997yq,donagi-1997-1,Friedman:1997ih}.
The computation of the vector-like matter spectrum  of these $U(N)$ bundles requires some technology from algebraic geometry which may be  slightly beyond the everyday needs in the physics literature. Before addressing the construction of string vacua of the above type in section 4, we therefore have to spend some time diving into the details of the spectral cover construction. In particular, it will be necessary to
generalise the  methods developed in \cite{Donagi:2004ia}.

For self-consistency of this paper we will begin section 2 by reviewing
 the construction of $\mu$-stable  bundles over elliptically fibered
Calabi-Yau manifolds via the method of spectral covers. 
We will also explain the method of bundle extensions, which allows
one to construct stable bundles of higher from lower rank ones. 
We then propose a new notion of stability, which should be relevant for 
vector bundles for heterotic strings and which includes, similarly to $\Pi$-stability
for D-branes, higher perturbative and non-perturbative corrections.
We call  bundles which are stable in this sense $\Lambda$-stable.
Finally, we recall the criterion stated in \cite{Donagi:2000zf} for stability of extensions of spectral cover bundles and prove it in appendix~\ref{proof}.

In section 3 we will partly review and partly newly derive the main technical tools for  the computation
of the various cohomology groups relevant for determining the massless
modes for the string compactifications of interest. By applying the Leray spectral sequence,
we will first recall that the cohomology of line bundles over the Calabi-Yau manifold
can be computed from line bundles over the base manifold of the elliptic fibration.
Moreover, we will explicitly show that the cohomology of the tensor product of two $U(N)$ bundles
localises on the intersection curve of the two  spectral covers and can be determined
by computing solely the cohomology of a certain line bundle  over this support curve.
This is in agreement with the special case considered in  \cite{Donagi:2004ia} that one of the bundles is a trivial line.
Therefore, eventually the entire computation is transformed into
computing the cohomology of line bundles over curves given by complete intersections
of two surfaces. These can be evaluated using  Koszul sequences.
With these results available we move forward and newly 
compute the cohomologies of the bundles $\bigwedge^2 V$
and ${\bf S}^2 V$. Note that the formula we derive differs from the one found
in \cite{Donagi:2004ia}.

Equipped with these powerful mathematical results, in section 4 we address
the construction of flipped $SU(5)$ heterotic string vacua by using vector bundles
with structure group $SU(4)\times U(1)$. After recalling the main ingredients of \cite{Blumenhagen:2006ux} we provide a new globally consistent supersymmetric three generation
example, for which the $U(4)$ bundle is defined as a stable extension of two $U(2)$ bundles. The model exhibits precisely one pair of GUT Higgs fields as required for GUT breaking down to the MSSM gauge group. The particle spectrum of the resulting $SU(3) \times SU(2) \times U(1)_Y$ vacuum is precisely that of the supersymmetric Standard Model spectrum with no extra vector-like matter but a number of additional electro-weak Higgs pairs. To the best of our knowledge, this is the first consistent flipped $SU(5)$ string model with these properties in the literature.

\section{Stable $U(n)$ bundles via spectral covers}

In sections \ref{seccy} - \ref{secdelPezzo} we  review the  construction  of $\mu$-stable $U(n)$ vector bundles over elliptically
fibered Calabi-Yau manifolds via the spectral cover method \cite{Friedman:1997yq,Friedman:1997ih,donagi-1997-1}. More information can also be found e.g. in \cite{Donagi:1999gc}. We then recall in section \ref{secext} the definition of vector bundles as non-trivial extensions of such spectral cover bundles. These parts are meant as a pedagogical introduction to this topic in order to make the present article  accessible to the non-expert reader and may safely be skipped by specialists. Section \ref{seclambda} analyses the stability concept appropriate for our applications, relegating a proof of the stability of our bundles to the appendix \ref{proof}.

\subsection{Elliptically fibered Calabi-Yau manifolds}
\label{seccy}

An elliptically fibered complex threefold $X$ is given by a complex two-surface $B$, the base space, together with an analytic map
\bea
\pi: X \rightarrow B,
\eea
where fibers over each point $b$ in the base
\bea
E_b=\pi^{-1}(b)
\eea
are elliptic curves.
Recall that an elliptic curve is a two-torus with a complex structure inducing an abelian group law. In particular it contains a distinguished point $p$ acting as the zero element in this group. 

\noindent We require the fibration $X$ to admit a global section $\sigma: B \rightarrow X$, assigning to every point in the base $b\in B$ the zero element $\sigma(b)=p \in E_b$ on the fiber. This section embeds the base as a sub-manifold into $X$
and we will often not distinguish between $B$ as a complex two-fold and $\sigma$ as its image in $X$.
The associated homology class in $H_4(X, {\mathbb Z})$ then intersects the fibre class precisely once. 
It will be useful to introduce also the class in $H^2(X, {\mathbb Z})$ Poincar\'e dual to the class of $\sigma$. In slight abuse of notation, it will also be referred to as $\sigma$. The respective meaning will hopefully always be clear from the context. Its cohomological self-intersection can be proven to be \cite{Friedman:1997yq} 
\bea
\sigma \cdot \sigma = - \sigma \cdot \pi^* c_1(B).
\eea 
Likewise, we introduce $F \in H^4(X, {\mathbb Z})$ as the Poincar\'e dual to the fibre class. 
 The fact that the base class intersects the class of the generic fibre once is reflected in the cohomological intersection form
\bea
\sigma \cdot F = 1.
\eea
This shows that $F$ is actually the Hodge dual to the two-form ${\sigma}$.
Now that we are at it, we state for later purposes the simple fact that the intersection form of the pull-back to $X$ of two classes $\a$ and $\b$ in $H^2(B, {\mathbb Z})$ is given by the pull-back of the intersection on $B$,
\bea
\pi^*(\a)\cdot \pi^*(\b ) = \pi^*( \a \cdot \b) = ( \a \cdot \b) \, F.
\eea

Let us now turn our attention to the elliptic fibre. Elliptic curves can be described as the hyperplane in ${\mathbb C \mathbb P}^2$ defined by the homogeneous Weierstrass equation
\bea\label{rW}
zy^2 = 4x^3-g_2xz^2-g_3z^3,
\eea
where $x,y,z$ are homogeneous coordinates on ${\mathbb C \mathbb P}^2$ and $g_2$ and $g_3$ define the complex structure.  
When we define a family of elliptic curves over the base, $x,y,z$ and $g_2$ and $g_3$ must be promoted to global sections of certain powers of some line bundle $\cal L $ on $B$.  The choice of this line bundle and the global sections $x,y,z$ will define the fibration.  

In order to promote equation (\ref{rW}) to a vanishing condition of a global section of a line bundle on $B$, we choose  $x,y,z$ to be sections of $\mathcal{L}^2$, $\mathcal{L}^3$ and ${\cal O}$ whereas  $g_2$ and $g_3$ appear as sections of $\mathcal{L}^4$ and $\mathcal{L}^6$, respectively.
If the fibration $X$ is to be Calabi-Yau, the first Chern class of the tangent bundle $T$ must vanish,
\bea
\label{CY}
c_1(X)=0.
\eea
As shown e.g. in \cite{Donagi:1999gc}, this implies $\mathcal{L}=K_B^{-1}$, where $K_B$ is the canonical bundle of the base space. It follows that $K_B^{-4}$ and $K_B^{-6}$ must have sections $g_2$ and $g_3$, respectively. The surfaces compatible with this condition are found to be del Pezzo, Hirzebruch, Enriques and blow-ups of Hirzebruch surfaces \cite{Morrison:1996pp}. Note, however, that the construction of stable holomorphic bundles on elliptically fibered three-folds does not hinge upon the Calabi-Yau property. In order to simplify the mathematical apparatus, we nonetheless assume (\ref{CY}) in the sequel.

Friedman-Morgen-Witten (FMW)  showed that on such spaces the Chern classes of the tangent bundle of the total space follow from the Chern classes of the base space. In particular, the second Chern class of the tangent bundle can be computed as
\bea
\label{tangentbundle}
c_2(X)=12\sigma \cdot  \pi^* c_1(B) + \left( 11c_1(B)^2+c_2(B) \right) F.
\eea

For later purposes let us recall that on $X$ there exists a holomorphic involution
$\tau$ acting solely on the fiber as $\tau:y\to -y$. The fixed point locus
of $\tau$ consists of two components. The first component is given by $x=z=0$ and arbitrary $y$, which is
nothing else than the section $\sigma$. The second component is defined by $y=0$ and is therefore
a triple cover of $B$, whose homology class was derived in \cite{Friedman:1997yq} 
as $3\sigma+ 3 c_1(B)$.
The homology class of the complete fixed point surface is therefore
\bea
         \sigma_{\tau}=4\sigma+3c_1(B),
\eea
where the factor $4\sigma$ reflects the four fixed points of the holomorphic involution $(-1)$
on $T^2$. 

\subsection{The spectral cover construction}
\label{secscc}

The basic idea of the spectral cover method is to first construct a $\mu$-stable $U(n)$ or $SU(n)$  bundle on the elliptic fibre over each point of the base, which is then extended over the whole manifold by gluing the data together suitably. Recall that in general, a $U(n)$ or $SU(n)$  bundle defines a rank $n$ complex vector bundle. Such a rank $n$ bundle over  an elliptic curve must, in order to satisfy the Hermitian Yang-Mills equation, be of degree zero.  More precisely, it can be shown to be isomorphic to the direct sum of $n$ complex line bundles
\bea
{\cal V}|_{E_b}=\mathcal{N}_1\oplus\ldots\oplus\mathcal{N}_n,
\eea
each of which has to be of zero degree. If $G=SU(n)$ as opposed to $U(n)$, ${\cal V}|_{E_b}$ must in addition be of trivial determinant, i.e. $\bigotimes_{i=1}^n {\cal N}_i = {\cal O}_{E_b}$. The zero degree condition on ${\cal N}_i$ implies that there exists for each  ${\cal N}_i$ a meromorphic section with precisely one zero at some $Q_i$ and a pole at $p$, i.e. ${\cal N}_i = {\cal O}_{E_b}(Q_i-p)$. Consequently, stable $(S)U(n)$ bundles on an elliptic curve are in one-to-one correspondence with the unordered $n$-tuple of points $Q_i$, and the reduction of $U(N)$ to $SU(n)$ is encoded in the additional requirement that $\sum_i(Q_i - p) = 0$ in the group law of the elliptic curve. 

Having understood the restriction of a rank $n$ bundle ${\cal V}$ to each elliptic fibre, we can now proceed to constructing the whole of ${\cal V}$. In intuitive terms, the above implies that over an elliptically fibered manifold a $U(n)$ vector bundle determines a set of $n$ points, varying over the base. More precisely, the bundle ${\cal V}$ over $X$ with the property that for a generic fiber $E_b$
\bea
\label{V1}
{\cal V}|_{E_b} = \bigoplus_{i=1}^n {\cal O}(Q_i-p)
\eea
 uniquely defines an $n$-fold cover $C$ of $B$, the spectral cover. It is defined by a projection
\bea
\pi_C: C \rightarrow B \quad\quad\quad\quad {\rm and} \quad\quad    C \cap E_b = \pi_C^{-1}(b) = \bigcup_i  \,Q_i. 
\eea
$C$, which is a hypersurface in $X$, can be conveniently described as the vanishing locus of some global section of the line bundle  ${\cal O}_X(n \sigma + \pi^* \eta)$. Here $\eta$ denotes some effective class in $H^2(B, {\mathbb Z})$. In particular, this implies that the  Poincar\'e dual two-form of $C$ is in 
\bea
\label{C-class}
[C] = n \sigma + \pi^* \eta \in H^2(X, {\mathbb Z})
\eea
Note that under the involution $\tau$ the class $[C]$ is invariant, while the spectral cover $C$ is in general not invariant.

Several distinct bundles over $X$ may well give rise to the same spectral cover $C$ since the latter only encodes the information about the restriction of ${\cal V}$ to the fibre $E_b$. To recover ${\cal V}$ from the spectral data we need to specify in addition how it varies over the base, i.e. ${\cal V}|_B$. As discussed in \cite{Friedman:1997yq}  this is uniquely accomplished by the so-called spectral line bundle ${\cal N}$ on $C$ with the property
\bea
\label{V2}
\pi_{C*} {\cal N} = {\cal V}|_B.
\eea

We can formalise these results by introducing the notion of the Poincar\'e line bundle ${\cal P}$.  
For this purpose, consider the fibre product $X \times _B X'$ as the set of pairs $(z_1,z_2) \in X \times X'$ with $\pi(z_1) = \pi(z_2)$. Furthermore we need to introduce   $\pi_1$ and  $\pi_2$ as the projections on the first and second factor, respectively. 
Moreover, $\sigma_1$ denotes the section $\sigma_1:B\to X \to X \times _B X'$ and $\sigma_2$ the
section $\sigma_2:B\to X' \to X \times _B X'$. 
Then  ${\cal P}$ is defined as the bundle over $X \times _B X'$ with the two properties
\bea
\label{Poinc}
{\cal P}|_{E_b \times x} \simeq {\cal P}|_{x \times E_b} \simeq {\cal O}_{E_b}(x-p), \quad\quad\quad
{\cal P} |_{\sigma_i}  = {\cal O}_{\sigma_i},\;i=1,2.
 \eea
Introducing the diagonal divisor $\Delta$, the first Chern class of the Poincar\'e line bundle is
\bea
     c_1({\cal P})=\Delta-\sigma_1-\sigma_2-c_1(B).
\eea
Note that $\Delta$ satisfies the relations
\bea
       \Delta^2=-\Delta\cdot c_1(B), \quad  \Delta\cdot \sigma_i=\sigma_1\cdot \sigma_2.
\eea
We will denote by  ${\cal P}_B$ the restriction of ${\cal P}$ to $X \times_B C$. 
Now by definition, $\pi_{1*}({\cal P}_B)|_x= \bigoplus_i{\cal O}_{E_{\pi(x)}}(Q_i-p)$, as is clear from the fact that   $ C \cap E_b  = \bigcup_i Q_i$ and the first property in (\ref{Poinc}). 
This remains true if we tensor ${\cal P}_B$ with $\pi_{2}^{\ast}({\cal N})$ for some line bundle ${\cal N}$ on $C$. After all, $\pi_2^*({\cal N})$ as a bundle on $X$  is trivial when restricted to the fibre $E_b$. On the other hand, ${\cal P}|_{\sigma \times_B X^{'}}$ is likewise trivial due to the second property in (\ref{Poinc}), and so
$(\pi_{1\ast}(\pi_2^{\ast}\mathcal{N}\otimes\mathcal{P_B}))|_B=\pi_{1\ast} (\pi_2^{\ast}\mathcal{N}\otimes\mathcal{P_B})|_{\sigma_2})$  is simply given by $\pi_{C*} {\cal N}$. In other words, the bundle

\bea
\label{defSCC}
{\cal V}=\pi_{1*}(\pi_2^{\ast}\mathcal{N}\otimes\mathcal{P}_B)
\eea

\noindent indeed exhibits the two defining properties (\ref{V1}) and ({\ref{V2}).
This establishes the definition of an $(S)U(n)$ bundle on the elliptically fibered Calabi-Yau threefold in terms of the spectral data $(C, {\cal N})$. We reiterate that we will only consider the case that the restriction of the bundle to the elliptic fibre is an $SU(n)$ bundle, i.e. that $C$ is as in (\ref{C-class}).  

The bundles constructed so far are  $\mu$-semi-stable on a generic elliptic fiber. It has been shown in \cite{Friedman:1997ih}, Theorem 7.1,  that an irreducible  spectral cover is a sufficient condition  in order to obtain a $\mu$-stable vector bundle.\footnote{In
  fact, the proof guarantees stability of the bundle with respect to an ample class, i.e. a K\"ahler class, $J=\epsilon \sigma +J_B$    such that the K\"ahler parameter of the
  fiber lies in a certain range near the boundary of the 
  K\"ahler cone, that is for
sufficiently small $\epsilon$. Since the value of $\epsilon$ is not
known, in all models involving the spectral cover constructions it is
  therefore a subtle issue if the region of stability overlaps with the
  perturbative regime, which is needed to have control over non-perturbative
  effects. In all examples which will be relevant for us, the constraints will leave us enough freedom to go to regions of the K\"ahler cone where $\epsilon$ is much smaller than $J_B$.} There are two simple  conditions  on the curve $\eta$ \cite{Donagi:2004ia} which ensure the existence of an irreducible spectral cover:
\begin{itemize}
\item The linear system $|\eta|$ is base-point free.
\item The class $\eta - nc_1(B)$ is  effective.
\end{itemize}
We will be more specific about their implications when it comes to a discussion of the properties of the base.

We now give the topological invariants of the bundle ${\cal V}$ defined by (\ref{defSCC}). The working horse for this computation is the Grothendieck-Riemann-Roch (GRR) theorem.
Applying this theorem to the projection  $\pi_1: X \times_B C \rightarrow X$ allows us to compute the Chern classes of ${\cal V}$
\bea
\label{GRR2}
\pi_{1*} \left( e^{c_1( {\cal N} \otimes {\cal P}_B)}\, {\rm Td}(X \times_B C ) \right) = {\rm ch}({\cal V}) {\rm Td}(X).
\eea
As discussed in  \cite{Friedman:1997yq}, this relates in particular $c_1({\cal N})$ and $c_1({\cal V})$ as
\bea
c_1(\mathcal{N})= \frac{1}{n}\,  \pi_C^{\ast} c_1({\cal V})|_B -\frac{1}{2}\,c_1(C)+\frac{1}{2}\,\pi_C^{\ast} c_1(B) + \gamma
\eea
in terms of the  cohomology class $\gamma$ satisfying
\bea
\pi_{C\ast}\gamma=0.
\eea
One can prove that $\gamma$ can in general be written as
\bea
\gamma=\lambda(n\sigma-\pi_C^{\ast}\eta+n\pi_C^{\ast}c_1(B)),
\eea
where $\lambda\in\mathbb{Q}$. Note furthermore that $c_1(C)$ is minus the
first Chern class of the canonical bundle $K_C = {\cal O}(C)$ on $C$,
i.e. $c_1(C) = -n \sigma - \pi_C^* \eta $.
We now parameterise $c_1({\cal V})$ by some element  $c_1(\zeta) \in H^2(B, {\mathbb Z})$ to be specified momentarily,
\bea
c_1({\cal V}) =\pi^*  c_1(\zeta).
\eea
Putting everything together, we have
\bea
\label{linebundle}
c_1(\mathcal{N})=n\left({\textstyle  \frac{1}{2}+\lambda }\right)\,\sigma +   
   \left({\textstyle \frac{1}{2} - \lambda}\right) \pi_C^\ast \eta + \left( {\textstyle \frac{1}{2}+ n
   \lambda} \right) \pi_C^\ast c_1(B) + {\textstyle \frac{1}{n}}\, \pi_C^\ast c_1(\zeta).
\eea
Since $c_1(\mathcal{N})$ and $c_1({\cal V})$ must be an integer class, not every value of $\lambda \in {\mathbb Q}$ and $c_1(\zeta) \in  H^2(B, {\mathbb Z})$ is allowed in the ansatz for $c_1({\cal V})$. 
Rather they are subject to the constraints
\bea
\label{Ninteger}
 n\left ({\textstyle \frac{1}{2}+  \,\lambda }\right)&\in& {\mathbb Z}, \nonumber \\
 \left({\textstyle \frac{1}{2}-\lambda}\right)\,\eta +\left({\textstyle n \lambda + \frac{1}{2}}\right)
 c_1(B) + 
      {\textstyle \frac{1}{n}}\, c_1(\zeta)   &\in& H^2(B, {\mathbb Z}),
\eea
but can otherwise be chosen arbitrarily. Note that if we are interested in $SU(n)$ bundles as e.g. in \cite{Friedman:1997yq}, then simply $c_1(\zeta)=0$ so that $c_1({\cal V}) = 0$. All other consistent choices yield $U(n)$ bundles. Allowing non-trivial values for $c_1({\cal V})$ was first considered in  \cite{Andreas:2004ja} and motivated by the relative Fourier-Mukai transform, but we will not invoke this picture here\footnote{To recover their expressions, simply set $c_1(\zeta) = \eta_E - \frac{n}{2} c_1(B)$ in the notation of \cite{Andreas:2004ja}.} . 
Further applications of the GRR theorem lead to the following expressions for the second and third Chern classes \cite{Friedman:1997yq,Curio:1998vu, Andreas:2004ja}
\bea
\label{Chern1}
\ch_2({\cal V}) &=& -\sigma  \cdot  \pi^* \eta + \left( {\textstyle\frac{1}{2 n}} c_1(\zeta)^2 - \omega \right) F, \nonumber \\
\ch_3({\cal V}) &=& \lambda \eta\cdot (\eta - n c_1(B))- {\textstyle \frac{1}{n}}\, c_1(\zeta)\cdot \eta,
\eea
where
\bea
\omega&=&-\frac{1}{24}  c_1(B)^2 (n^3-n) + \frac{1}{2} \left( \lambda^2 - \frac{1}{4}\right) n\eta\cdot (\eta-nc_1(B)).
\eea
 Note that $\ch_3(V)$ has already been integrated over the fiber.

As we emphasized several times, this kind of construction only gives bundles with trivial first Chern class as restricted to the elliptic fibres. To be more general, we can however twist the bundle ${\cal V}$ defined via the spectral cover construction 
with an additional line bundle $\mathcal{Q}$ on $X$ with  \cite{Blumenhagen:2005zg}
\bea
\label{c1Q}
c_1(\mathcal{Q})=q\sigma + \pi^*(c_1(\zeta_Q)),
\eea
where $ \pi^*(c_1(\zeta_Q)) \in H^2(X, \mathbb Z)$. The resulting $U(n)$ bundle
\bea
V={\cal V}\otimes\mathcal{Q}
\eea
is $\mu$-stable precisely if the original bundle ${\cal V}$ is. The Chern classes for $V$ are straightforwardly computed from  the ones of ${\cal V}$  and from $c_1({\cal Q})$. Note that the contribution form $\pi^*(c_1(\zeta_Q))$ can always  be absorbed into an additive shift of $c_1(\zeta)$ by $ n c_1(\zeta_Q)$.
We will not make use of $U(n)$ bundles with $q \neq 0$ in this article. The above Chern characters are therefore sufficient for our purposes.

\subsection{Del Pezzo surfaces}
\label{secdelPezzo}

As alluded to already, the Calabi-Yau condition imposes severe constraints on which complex two-surfaces are eligible as base manifolds of our elliptic fibration.
Among the possibilities classified in \cite{Morrison:1996pp} we can choose as the base manifold  one of the
 del Pezzo surfaces dP$_r$ with $r=0,\ldots,9$. The surface dP$_r$ is defined by blowing up
$r$ points in generic position on $\IP_2$. This means that $H^2(\rm{dP}_r,\mathbb{Z})$ is generated by the $r+1$ 
elements $l,E_1,\ldots,E_r$, where $l$ is the hyperplane class inherited from 
$\IP_2$ and the $E_m$ denote the $r$  exceptional cycles introduced
by the blow-ups. 
The intersection form can be computed as
\bea
\label{Intform}
      l\cdot l=1, \quad l\cdot E_m=0, \quad E_m\cdot E_n=-\delta_{m,n}.
\eea
The first equation follows from the fact that two representatives of the class $l$ define two complex lines in generic position which clearly intersect precisely once. The self-intersection for the blow-ups is the usual one for exceptional cycles. Furthermore, a complex line in generic position does not pass through any of the blown-ups, thus $l\cdot E_m=0$.

The Chern classes read
\bea
   c_1(dP_r)=3l-\sum_{m=1}^r  E_m, \quad\quad c_2(dP_r)=3+r.
\eea

We recognize the part involving $l$ as  the first Chern class  of the parent ${\mathbb P}_2$.
For the second Chern class of the elliptic threefold $X$ we obtain, applying (\ref{tangentbundle}),
\bea
   c_2(X)=12 \sigma c_1(B) +(102-10r)\, F.
\eea
Now for a vector bundle $V_i$ we can expand $\eta_i$ and $c_1(\zeta_i)$ in a cohomological basis
\bea
   \eta_i&=&\eta_i^{(0)}\, l + \sum_{m=1}^r \eta_i^{(m)}\, E_m \equiv (\eta_i^{(0)},\eta_i^{(1)}, \ldots,\eta_i^{(r)})  \nonumber \\
   c_1(\zeta_i)&=&\zeta_i^{(0)}\, l + \sum_{m=1}^r  \zeta_i^{(m)}\, E_m \equiv (\zeta_i^{(0)},\zeta_i^{(1)}, \ldots,\zeta_i^{(r)}).
\eea    
As mentioned before we have to require for stability that $|\eta|$ is effective
and that  $\eta-n\, c_1(B)$ is effective. Fortunately, the generating system for the cone of effective curves on dP$_r$ has been
given in \cite{Demazure} and we list the reformulated result of~\cite{Donagi:2004ia}
in Table~\ref{TdPrGenerators} for
completeness. Recall that a general effective class can be expanded into a linear combination of these Mori cone generators with non-negative integer coefficients.
\begin{table}[htb]
\renewcommand{\arraystretch}{1.5}
\begin{center}
\begin{tabular}{|c||c|c|}
\hline
\hline
$r$ & Generators & $\#$  \\
\hline \hline
1 & $E_1$, $l-E_1$  & 2 \\\hline
2 & $E_i$, $l-E_1-E_2$  & 3 \\\hline
3 & $E_i$, $l-E_i-E_j$  & 6 \\\hline  
4 & $E_i$, $l-E_i-E_j$  & 10 \\\hline  
5 & $E_i$, $l-E_i-E_j$, $2l-E_1-E_2-E_3-E_4-E_5$ & 16  \\\hline    
6 & $E_i$, $l-E_i-E_j$, $2l-E_i-E_j-E_k-E_l-E_m$ & 27  \\\hline        
7 & $E_i$, $l-E_i-E_j$, $2l-E_i-E_j-E_k-E_l-E_m$, & \\
 &  $3l-2E_i-E_j-E_k-E_l-E_m-E_n-E_o$ & 56  \\\hline       
8 & $E_i$, $l-E_i-E_j$, $2l-E_i-E_j-E_k-E_l-E_m$, & \\
 &  $3l-2E_i-E_j-E_k-E_l-E_m-E_n-E_o$, & \\  
 & $4l-2(E_i+E_j+E_k)-\sum_{i=1}^5 E_{m_i}$, & \\
 & $5l-2\sum_{i=1}^6E_{m_i}-E_k-E_l$, $6l-3E_i-2 \sum_{i=1}^7 E_{m_i}$ & 240  \\\hline   
9 & $f=3-\sum_{i=1}^9 E_i$, and $\{y_a\}$ with $y_a^2=-1$, $y_a \cdot f=1$ & $\infty$
\\
\hline
\end{tabular}
\caption{\small Generators for the Mori cone of each dP$_r$, $r=1,\ldots,9$. All indices $i,j,\ldots \in \{1,\ldots,r\}$ 
in the table are distinct. The effective classes can be written as linear combinations of the generators with integer 
non-negative coefficients. }
\label{TdPrGenerators}
\end{center}
\end{table}

Moreover, $|\eta|$ is known to be base point free if $\eta\cdot E\ge 0$
for every curve $E$ with $E^2=-1$ and $E\cdot c_1(B)=1$.  Such curves
are precisely given by the generators of the Mori cone listed
in Table~\ref{TdPrGenerators}.

\subsection{More bundles from extensions}
\label{secext}

The $U(n)$ bundles constructed in the previous section can serve as the building block for a more general construction of vector bundles known as the extension method. Physically, the idea is to start with the direct sum of two bundles, $V_1 \oplus V_2$ and deform it into a new, stable bundle $V$. More abstractly, if such a deformation is possible, the resulting bundle $V$, the extension of $V_2$ by $V_1$, fits into the short exact sequence
\bea
\label{ext_gen}
0 \rightarrow V_1 \rightarrow V \rightarrow V_2 \rightarrow 0.
\eea
The possible deformations of $V_1 \oplus V_2$ which still fit into the exact sequence (\ref{ext_gen}) are classified by the extension group ${\rm Ext}^*_X(V_2, V_1)$. In our case, since $V_1$ and $V_2$ are vector bundles and not merely coherent sheaves, the extension group is actually given by the cohomology groups $H^*(X, V_1 \otimes V_2^\vee)$. The above extension $V$ can be chosen non-split, i.e. V is a proper deformation of $V_1 \oplus V_2$, precisely if
\bea
 H^1(X, V_1 \otimes V_2^\vee) \neq 0.
\eea

\noindent The total Chern character of the extension bundle $V$ follows from the ones of $V_1$ and $V_2$ as
\bea
\ch(V) = \ch(V_1) + \ch(V_2).
\eea

\noindent The cohomology groups of $V$, $H^*(X, V)$, can in principle be computed from $H^*(X, V_1)$ and $H^*(X, V_2)$ by exploiting the standard fact that a short exact sequence induces a long exact sequence in cohomology.

For later use we note furthermore that the exact sequence (\ref{ext_gen}) remains exact upon tensoring each element appearing in it by a line bundle $L$, i.e. the sequence
\bea
\label{ext_L}
0 \rightarrow V_1 \otimes L \rightarrow V \otimes L \rightarrow V_2 \otimes L\rightarrow 0
\eea
is exact precisely if (\ref{ext_gen}) is.
This will allow us to obtain the cohomology groups of $V \otimes L$ from  $H^*(X, V_i \otimes L)$ by invoking the long exact sequence in cohomology induced by (\ref{ext_L}).

\subsection{Comments on $\mu$- and $\Lambda$-stability}
\label{seclambda}

At string tree level, for a heterotic compactification to preserve
supersymmetry,  the vector bundle must be holomorphic and its field strength has to satisfy the Hermitian Yang-Mills (HYM) equation $g^{a\overline b}\, F_{a \overline b}=0$. The latter is most conveniently rewritten in its dual version
\bea
            \star_6 \bigl[ J\wedge J \wedge F^{ab}_i \bigr]=0,
\eea
where $a,b$ are gauge indices and  $i=1,2$ distinguishes the two $E_8$ factors. 
A solution to this equation exists if the vector bundle is $\mu-$stable and
obeys the Donaldson-Uhlenbeck-Yau (DUY) condition
\bea
\label{DUY}
            \int_X J \wedge J \wedge c_1(V)=0.
\eea
Recall that a vector bundle $V$ is called $\mu$-stable if 
each subsheaf ${\cal F}$ of rank smaller than the rank of $V$  satisfies
 $\mu({\cal F}) < \mu(V)$, where the $\mu$-slope $\mu({\cal F})$ for a sheaf ${\cal F}$ 
with respect to  the K\"ahler form $J$ of the manifold $X$ is defined as 
\bea
\mu({\cal F}) = \frac{1}{{\rm rk}{\cal F}} \int_X J \wedge J \wedge c_1({\cal F}).
\eea

As has been shown in \cite{Blumenhagen:2005ga} by analysing the D-term supersymmetry conditions in the effective four-dimensional ${\cal N}=1$  supergravity, for $U(N)$ bundles there exists a one-loop
correction to the DUY equation. Following the same logic
which has lead to the DUY theorem, it is natural to  conjecture that this is due to  
a corresponding stringy one-loop correction to the HYM equation of the form
\bea
\label{HYMloop}
     && \star_6 \Biggl[ J\wedge J \wedge F^{ab}_i
      - {\frac{\ell_s^4}{4(2\pi)^2}}\,   e^{2\phi_{10}}\,\, F^{ab}_i \wedge 
        \left({\rm tr}_{E_{8i}}(F_i\wedge F_i) -{\frac{1}{2}}
         {\rm tr}(R\wedge R)\right)  \\
      &&\phantom{aaaaaaaaaa} + \ell_s^4 e^{2\phi_{10}}\, \sum_a N_a \left( {\frac{1}{2}}\mp \lambda_a\right)^2
          F^{ab}_i\wedge \ov\gamma_a\ \Biggr]+\ ({\rm n.p.\ terms})=0. \nonumber
\eea
Here $\ov\gamma_a$ denotes
the Poincar\'e dual four-form of the two-cycles wrapped by five-branes which may or may not be present in the concrete vacuum under consideration. 
The positions of the five-branes are parametrized by $-1/2\le \lambda_a \le 1/2$
and the minus sign in the last term in (\ref{HYMloop}) is for the first $E_8$ and the plus sign for the second. More information can be found in \cite{Blumenhagen:2006ux}. 
Non-renormalisation theorems for D-terms in supersymmetric theories imply that there
are no higher loop corrections, but as indicated there can be non-perturbative ones.

In view of the above quantum corrections to the HYM equation it is clear that the stability concept relevant for finding solutions to (\ref{HYMloop}) likewise has to be modified. 
As with $\Pi$-stability for B-type D-branes \cite{Douglas:2000ah}, 
in the $E_8\times E_8$ heterotic string this new notion of stability would
correct the tree-level concept  of $\mu$-stability \footnote{$\Pi$-stability is meant to be
the correct notion of stability for B-type D-branes in the limit $g_s=0$ and to all orders in $\alpha'$.
By S-duality one is tempted to introduce the corresponding stability for heterotic
bundles in the limit $g_s\to \infty$, $\alpha'\to0$ with $\alpha' g_s=const.$}.

If we were not dealing with the zero-slope equation (\ref{HYMloop}), but instead allowed for some unspecified term ${\rm const.} \,  \times {\rm vol.} \, \,  {\rm id}$ on the righthand side, the situation would be very similar to the perturbative deformation of the HYM equation as encountered in the context of Gieseker stability \cite{Leung}.
More precisely,
we would like to conjecture that this more general, complete loop and non-perturbative 
corrected HYM equation,

\bea
\label{HYMloop2}
     && \star_6 \Biggl[ J\wedge J \wedge F^{ab}_i
      - {\frac{\ell_s^4}{4(2\pi)^2}}\,   e^{2\phi_{10}}\,\, F^{ab}_i \wedge 
        \left({\rm tr}_{E_{8i}}(F_i\wedge F_i) -{\frac{1}{2}}
         {\rm tr}(R\wedge R)\right) + \\
      &&\phantom{aa} \ell_s^4 e^{2\phi_{10}}\, \sum_a N_a \left( {\frac{1}{2}}\mp \lambda_a\right)^2
          F^{ab}_i\wedge \overline\gamma_a\ \Biggr]+\ ({\rm n.p.\ terms}) 
          = {\rm const}. \, \times {\rm vol.} \, \, {\rm id}^{ab}, \nonumber
\eea
has a solution if the bundle is stable with respect to a corrected slope
$\Lambda({\cal F})=\arg Z({\cal F})$
with the central charge
\bea
\label{Lstable}
   Z({\cal F})&=&{\frac{1}{ 2\pi g_s \ell_s^6}} {\rm Tr} \int_{X} e^J\, (1+2\pi i\alpha' g_s {\cal F} ) 
             \Biggl[
                 1-{\frac{\ell_s^4 g_s^2}{2}} \biggl( \frac{1}{ 4(2\pi)^2} 
              \biggl( {\rm tr}_{E_{8i}}(F_i\wedge F_i) \nonumber\\
	  && -\frac{1}{2} {\rm tr}(R\wedge R) \biggr) -\sum_a N_a \left( \frac{1}{2}\mp \lambda_a\right)^2
                 \ov\gamma_a \biggr) \Biggr] +\ ({\rm n.p.\ terms}).
\eea

We call such a bundle  $\Lambda$-stable and,
neglecting the unknown non-perturbative corrections in (\ref{Lstable}),
we call it  $\lambda$-stable. 
The reasoning behind this statement is that the tree-level part on the lefthand side of (\ref{HYMloop2}) can be tuned to dominate arbitrarily over the quantum corrections by choosing the expansion parameter $g_s$ correspondingly small. For more information we refer to \cite{Thesis}. 
This is no longer possible as soon as we insist that, a forteriori, (\ref{HYMloop}) is satisfied, which induces in addition $\Lambda(V) =0$. After all, we are now cancelling the tree-level and the higher order parts against each other.
A more refined analysis of the general quantum corrected stability concept is therefore desirable.

Luckily, for $SU(N)$ bundles and the particular type of $U(N)$ bundles
treated in this paper a simplification occurs since we will be interested in special bundles defining a heterotic compactification with gauge group flipped $SU(5) \times U(1)_X$ and $SU(3) \times SU(2) \times U(1)_Y$.
As has been shown in \cite{Blumenhagen:2006ux}, the same conditions on the bundles rendering the $U(1)_X$ and
$U(1)_Y$ massless imply that 
\bea
\label{l=m}
              \lambda(V)=\mu(V)=0.
\eea
Clearly, this also holds trivially for all $SU(N)$ bundles.
Whenever (\ref{l=m}) applies, the above arguments imply that $\lambda$-stability guarantees the existence of a solution to (\ref{HYMloop}). Moreover, for a $\mu$-stable bundle $V$, we can immediately conclude in this case that it
is also $\lambda$-stable for sufficiently small string coupling $g_s$,
 as for the finite number of subsheaves we can tune $g_s$ such that 
\bea
              \lambda({\cal F})=\mu({\cal F})+ O(g_s^2) < \mu(V)=\lambda(V).
\eea
This is the reason why it is safe for us to work with $\mu$-stable $U(N)$ bundles,
about which much more is known.

We now collect the conditions for $\mu$-stability of our extension bundles $V$ as defined in (\ref{ext_gen}). Since $V_1$ and $V_2$ are both constructed via irreducible spectral covers, they are guaranteed to be $\mu-$stable with respect to a suitable polarisation, as reviewed in  section \ref{secscc}.
A necessary condition for the extension (\ref{ext_gen}) to yield again a \emph{stable} vector bundle is clearly that it be non-split and that $\mu(V_1) < \mu(V) =0$. Otherwise $V_1$ would be a subbundle of $V$ with slope not smaller than that of $V$. It was stated in \cite{Donagi:2000zf} that this condition is also sufficient.
 To our knowledge, no proof of this assertion, upon which various  models in the literature rely, has been given \footnote{This was true until the very recent preprint \cite{AndreasCurio}, which appeared after our independent analysis on this point had been completed.}. Appendix \ref{proof} contains a detailed proof of this statement. More precisely  we will show there that 
 \bea
\label{stab_ext1}
{\mbox {V is}}\,  \mu{\rm -stable \, \, w.r.t.}\, J \Longleftrightarrow   H^1(X, V_1 \otimes V_2^\vee) \neq 0 \,\, \, {\rm and} \, \, \, \mu(V_1) < \mu(V).\eea

The condition (\ref{stab_ext1}) can be read as a constraint on the K\"ahler form $J$ of the manifold $X$ and has to be satisfied inside the K\"ahler cone such that also $V_1$ and $V_2$ are simultaneously stable with respect to it.

\section{Computation of cohomology classes}

Let us now come to the main technical section of this paper, where we partly review and
partly newly derive the mathematical formalism for the computation of the relevant
vector bundle cohomology classes. The computation of $H^i(X,V)$ has already
been described in very much detail in \cite{Donagi:2004ia}. Here, we instead compute the more
general classes $H^i(X,V_a\otimes V_b)$ and show, using the Leray spectral sequence,
that they are localised on the curve $C_a\cap C_b$.  The cohomology classes $H^i(X,\bigwedge^2 V)$ 
were also covered in \cite{Donagi:2004ia}, but our more physically inspired  approach
gives a deviating  result, which however is consistent with the Riemann-Roch-Hirzebruch 
theorem on the support curve. We also provide the computation
of the cohomology classes $H^i(X,{\bf S}^2 V)$.\footnote{We thank Stefano Guerra for pointing out to us that his upcoming work \cite{Guerra} analyses related questions.}

An important tool for the computation of the cohomology of vector bundles on elliptic fibrations is the Leray spectral sequence.
More generally, for any fibration $\pi:X\to B$, the Leray spectral sequence relates the cohomology of any bundle $V$ on $X$ to the cohomology of certain sheaves on the base $B$. These sheaves are called higher direct image sheaves $R^i \pi_* V$ and are defined by
\bea
R^i \pi_* V(U) = H^i(\pi^{-1}(U), V|_{\pi^{-1}(U)})
\eea
for any open set $U\subset B$. In particular, observe that for any point $b \in B$
\bea
\label{rightder}
R^i \pi_* V|_b = H^i(f_b, V|_{f_b}),
\eea
that is, the higher image sheaf captures the cohomology of $V$ along the fibers $f_b$ of $\pi$. In the case of an elliptic fibration, only $R^0\pi_*$ and $R^1\pi_*$ are non-zero and the Leray sequence degenerates to the 
long exact sequence
\begin{equation}
  \label{eq:W1les}
  \vcenter{\xymatrix@R=10pt@M=4pt@H+=22pt{
      0 \ar[r] & H^1(B,\pi_\star V)  \ar[r] &
       H^1(X, V)  \ar[r] &
      H^0(B, R^1 \pi_\star V)  
      \ar`[rd]^<>(0.5){}`[l]`[dlll]`[d][dll] & 
      \\
      & H^2(B,\pi_\star V) \ar[r] &
       H^2(X, V)   \ar[r] &
      H^1(B, R^1 \pi_\star V) \ar[r] &
      0
      \,
    }}
\end{equation}
together with
\begin{equation}
H^0(X, V)=H^0(B,\pi_\star V),\;\;\; H^3(X, V)=H^2(B,R^1 \pi_\star V).
\end{equation}
In addition, Serre duality on the one-dimensional fiber implies relative duality, i.e.
\bea
           \left(R^1 \pi_\star V \right)^\vee &=& \pi_\star (V^\vee \otimes K_X\otimes \pi^\star K_B^\vee) .
\eea
Another useful relation is the projection formula
\bea
      R^q\pi_\star( V\otimes \pi^\star {\cal F})=R^q\pi_\star( V)\otimes {\cal F},
\eea
for any vector bundle $\cal F$  on $B$.
To obtain information about the Chern classes  of the higher image sheaves one can use  the  Grothendieck-Riemann-Roch theorem
\bea
\label{GRR}
       \pi_{\star}\biggl( {\rm ch}(V) \, 
         {\rm Td}(X)\biggr)=         {\rm ch(\pi_! V)}\, {\rm Td}(B)
\eea
with 
\bea
             \pi_! V=\sum_{i=1}^d (-1)^i\, R^i \pi_* V
\eea
for fiber dimension $d$.
In (\ref{GRR}) the push-forward $\pi_{\star}$ of a  form is 
defined as integration over the fiber. The Todd classes are defined in terms of Chern classes as
\bea
   {\rm Td}(X)=1+{c_1(X)\over 2} +{c_2(X)+ c_1^2(X)\over 12} + {c_1(X)\, c_2(X)\over 24} +\ldots.
\eea
and simplify considerably for Calabi-Yau manifolds with $c_1(X)=0$.

\subsection{Cohomology classes  $H^i(X,{\cal L})$} 
\label{sec_H(L}

In order to compute the cohomology classes of vector bundles $V$ on $X$, we need to know how to compute the cohomology classes of any line bundle $\cal L$ on $X$.  Henceforth, $X$ will be a generic elliptic fibration over a complex two dimensional surface $B$ with zero section $\sigma$. Then any line bundle on $X$ will be of the form
\begin{equation}
\cL= {\cal O}_X(n \sigma) \otimes \pi^{\star} L
\end{equation}
for some line bundle $L$ on $B$. Applying the projection formula gives
\bea
        \pi_\star \cL=\pi_\star {\cal O}_X(n \sigma)\otimes  L, \quad\quad
        R^1\pi_\star \cL=R^1\pi_\star {\cal O}_X(n \sigma)\otimes  L.
\eea
The higher image sheaves of $\cO_X(n \sigma)$ for any $n$ are given by \cite{Donagi:2004ia},
\bea\label{pi-nsig}
\pi_*(\cO_X(n \sigma)) &=& \left\{
\ba{lll}
\cO_B \oplus \cO_B(-2 c_1(B)) \oplus \ldots \oplus \cO_B(-n c_1(B))
& n \geqslant 0 \\
0 & n < 0
\ea
\right. \nn \\
R^1\pi_*(\cO_X(n \sigma)) &=& \left\{
\ba{lll}
0 & n > 0 \\
\cO_B((-n-1)c_1(B)) \oplus \ldots \\
\ldots\oplus \cO_B(c_1(B)) 
\oplus \cO_B(- c_1(B))
&  n \leqslant 0. \\
\ea
\right. \nn \\
\eea
Therefore, in order to apply the Leray spectral sequence, 
one merely  has to determine  the cohomology classes
$H^i(B,L)$ 
of general line bundles over the base $B$. In our case $B$  is a del-Pezzo surface $dP_r$ and we  relegate our derivation of $H^i(dP_r ,L)$ to appendix \ref{app_dPr}.

\subsection{Cohomology classes  $H^i(X,V_a\otimes V_b)$}
\label{sec_VaVb}
In this section we will show how to compute the cohomology of $V_a\otimes V_b$,  where both $V_a$ and $V_b$ are vector bundles on $X$ which admit an irreducible spectral cover.

These cohomology classes are necessary to compute the cohomology of  the  vector bundles constructed via extensions as described in section \ref{secext}. In addition, they provide a natural setup for computing $H^i(X,\bigwedge^2 V)$ and $H^i(X,{\bf S}^2 V)$ for vector bundles  $V$ given by the   spectral cover construction.
Note that the cohomology groups $H^i(X,V_a\otimes V_b)={\rm {Ext}}^i_X(V_a^\vee,V_b)$ also  appear naturally in the $SO(32)$ heterotic  and the S-dual Type I string where they count matter
fields in bifundamental representations of an $U(N_a)\times U(N_b)$ gauge group \cite{Blumenhagen:2005pm, Blumenhagen:2005zg,Blumenhagen:2005zh}.

To begin with, it is useful to review the results of \cite{Donagi:2004ia} for the computation of the cohomology of a vector bundle $V$ with vanishing first Chern class given by an irreducible spectral cover $C$
and a globally defined line bundle $\cal N$. In oder to use the Leray sequence for $\pi: X \to B$ one must find $\pi_* V$ and $R^1\pi_*V$. Recall from section \ref{secscc} that the restriction of $V $ to a generic fiber $f_b$ for $b \in B$ is given by
\begin{equation}
\label{rVf}
V|_{f_b} = \oplus_{i=1}^{rk(V)} {\cal O}_{f_b}(Q_i - p),
\end{equation}
where $p$ and the points ${Q_i}$ denote the intersection of the fiber $f_b$ with $\sigma$ and  the spectral cover respectively. All of these points  are  disjoint under our assumption of an irreducible spectral cover. Therefore, $\pi_*V|_b=H^0(f_b, V|_{f_b})$ vanishes for a generic fiber. In addition, for any vector bundle $V$, $\pi_* V$ is torsion free and hence we find that $\pi_* V $ vanishes identically. It also follows from these considerations that $R^1\pi_* V$ is a sheaf on $B$ supported on the curve $c= {C} \cap \sigma$ where we identify $B\cong \sigma$. It was shown in \cite{Donagi:2004ia} that
\begin{equation}
R^1\pi_* V= {\cal N} \otimes K_B |_c.
\end{equation}
Applying these results to the Leray spectral sequence determines
\begin{equation}
H^0(X,V)=H^3(X,V)=0 ,\;\;H^i(X,V)=H^{i-1}(c,{\cal N} \otimes K_B |_c), \;i=1,2.
\end{equation}
In particular, it follows that
\begin{equation}
-\chi(X,V)=\chi(c,{\cal N}\otimes K_B),
\end{equation}
a result which can be easily checked with the help of the numerical expressions for $c_3(V)$, ${\cal N}$ and $C$ of section \ref{secscc}.

Consider now two vector bundles $V_a$ and $V_b$ with structure groups 
$U(n_a)$ and $U(n_b)$ 
given by the spectral cover construction. 
The  two irreducible spectral covers  $C_a$ and $C_b$ are in the linear system
\bea
    C_a \in |n_a\, \sigma + \pi^* \eta_a|,\;\; \quad C_b\in |n_b\, \sigma + \pi^* \eta_b|
\eea
and the spectral line bundles ${\cal N}_a$ and ${\cal N}_b$ are defined as in
(\ref{linebundle}). Note that the case  $V_a={\cal O}_X$ is included by choosing  $C_a=\sigma$ and ${\cal N}_a={\cal O}_X$. 
It follows from the discussion above that the basic strategy to compute the cohomology of $V_a\otimes V_b$ is to compute its spectral data. In particular, we need to find $H^i(c_{ab},{\cal N}_{ab} \otimes K_B|_{c_{ab}}),\,i=1,2$ where ${\cal N}_{ab}$ is the spectral line bundle, or more generally, a rank one sheaf corresponding to $V_a\otimes V_b$ and $c_{ab}$ is the intersection of the spectral cover of $V_a\otimes V_b$ with the zero section.

Before we attempt to compute ${\cal N}_{ab}|_{c_{ab}}$, it is instructive to consider the F-theoretic realization. There the chiral matter is defined by ${\rm {Ext}}^i_X(V^\vee_a,V_b)$, which is expected to be localized on the intersection of the  spectral cover for $V_a^\vee$ and $V_b$, i.e. on the intersection of the two stacks of D7-branes. It follows from (\ref{rVf}) that the spectral cover for $V_a^\vee$ is $\tau C_a$, with $\tau$ being the involution defined at the end of section \ref{seccy}. Generically, $\tau C_a \cap C_b$ is a smooth curve, denoted by $D$ in the sequel, whose cohomology class is  
\bea
   [D] \equiv  [\tau C_a\cap C_b] &=&\pi^* \left(-n_a n_b\, c_1(B) + n_a\, \eta_b + n_b \eta_a \right) \sigma + (\eta_a \eta_b )\, F \nonumber \\
                  &=& \pi^* [{c}_{ab}]\, \sigma + a_F \, F.
\eea
By $c_{ab}$ we denote the projection of the curve on the base. 
It can be shown using the techniques of \cite{Donagi:2004ia} that the so-defined class $[c_{ab}]$ is indeed the class of the intersection of the spectral cover of $V_a \otimes V_b$ with the zero section $\sigma$.

To find the points $b \in B$ which are contained in $c_{ab}$ consider the restriction of $V_a \otimes V_b$ to a generic fiber $f_b$
\bea
V_a \otimes V_b |_{f_b}=V_a|_{f_b} \otimes V_b |_{f_b}=\oplus_{i=1}^{rk(V_a)} {\cal O}_{f_b}(Q^a_i - p) \otimes \oplus_{j=1}^{rk(V_b)} {\cal O}_{f_b}(Q^b_j - p).
\eea
That is, the intersection points of the spectral cover of $V_a\otimes V_b$ with the fibre $f_b$ are given by the set $\{Q^a_i + Q^b_j\}_{ij}$ and the points $b \in c_{ab}$ are the subset thereof defined by $Q^a_i$ and  $Q^b_j$ in the fiber $f_b$ such that
$$
Q^a_i + Q^b_j =0.
$$
Here we take addition in the group  law of $f_b$. 
$D$ and $c_{ab}$ are related via the surjective map
$$
\pi_D: \tau C_a\cap C_b \to c_{ab},
$$
which is generically one-to-one. As noted above, $D$ is generically smooth, while it is far from obvious that $c_{ab}$ is.

It follows from this discussion that $\pi_*(V_a\otimes V_b)$ vanishes identically. Using the Leray spectral sequence and Serre duality this implies that
$$
H^0(X,V_a\otimes V_b)=H^3(X,V_a\otimes V_b)=0.
$$
In addition, $R^1\pi_*(V_a\otimes V_b)$ has support along the curve $c_{ab}$ and is given by $R^1\pi_*(V_a\otimes V_b)={\cal N}_{ab}\otimes K_B|_{c_{ab}}$.

To derive an expression  for ${\cal N}_{ab}|_{c_{ab}}$, recall from section~\ref{secscc} that
\bea
V|_B = \pi_{C*} {\cal N}.
\eea
Hence, at a generic point $b\in B$, we find
\bea\label{rVb}
V|_b=H^0(\pi_{C}^{-1}(b),{\cal N}|_{\pi_{C}^{-1}(b})=\oplus_i ({\cal N}|_{Q_i}).
\eea
That implies for the tensor product $V_a \otimes V_b$
\bea
V_a \otimes V_b|_b = \oplus_{ij} ({\cal N}_{ab}|_{Q^a_i+Q^b_j}) = \oplus_{ij} ({\cal N}_a|_{Q^a_i} \otimes {\cal N}_b|_{Q^b_j}).
\eea
The first and second equality follow from the application of formula  (\ref{rVb}) to $V_a \otimes V_b$ and to $V_a$ and $V_b$, respectively. Hence we see that the fiber of ${\cal N}_{ab}$ at $Q_i+Q_j$ is the tensor product of the fiber of ${\cal N}_a$ at $Q^a_i$ with the fiber of ${\cal N}_b$ at $Q^b_j$.

Let us assume that $b \in c_{ab}$. Following the discussion above, this implies that there are two points $Q^a_i$ and $Q^b_i$ obeying $Q^b_i = -  Q^a_j$. Replacing ${\cal N}_a$ with $\tau^*{\cal N}_a$ we find
\bea
{\cal N}_{ab}|_{Q^a_i+Q^b_j}=\tau^*{\cal N}_a|_{-Q^a_i} \otimes {\cal N}_b|_{Q^b_j}.
\eea
This description is certainly correct for a generic point $b\in c_{ab}$. Therefore, a natural conjecture for the spectral rank one sheaf of $V_a\otimes V_b$ restricted to $c_{ab}$ is
\bea
{\cal N}_{ab}|_{c_{ab}}=\pi_{*}(\tau^* {\cal N}_a \otimes {\cal N}_b|_{\tau C_a\cap C_b}).
\eea
In particular,  using the finiteness of $\pi_D$, we find for the cohomology of $V_a \otimes V_b$
\bea
\label{homolh}
     && H^{1}(X,V_a\otimes V_b)=H^0(\tau C_a\cap C_b,\, \tau^*{\cal N}_{a}\otimes {\cal N}_{b} \otimes K_B),   \\
    && H^{2}(X,V_a\otimes V_b)=H^1(\tau C_a\cap C_b,\,\tau^* {\cal N}_{a}\otimes {\cal N}_{b} \otimes K_B).  \nonumber 
\eea
Observe that for simply connected Calabi-Yau threefolds the Picard group is discrete. Therefore, spectral line bundles ${\cal N}_a$ with $\tau^*c_1({\cal N}_a)=c_1({\cal N}_a)$ satisfy  $\tau^*{\cal N}_a={\cal N}_a$. Note that the spectral line bundles appearing in our applications are just of this type.

In the sequel we will give a numerical proof of these results using general properties of the Fourier-Mukai transform. In particular, we will prove that 
$$
H^{i}(X,V_a\otimes V_b)=H^{i-1}(D, {\cal L}),\;i=1,2
$$
for some rank one sheaf $\cal L$ with 
$$
c_1({\cal L})=c_1({\cal N}_{a}\otimes {\cal N}_{b} \otimes K_B))|_D.
$$
To begin with, note that the projection formula allows us to write
\bea
     V_a\otimes V_b =\pi_{1\star}\left( {\cal P}_B\otimes \pi^\star_2 {\cal N}_b \otimes \pi_1^* V_a \right).
\eea
Hence we have the two maps 
\begin{equation}
  \label{Fouriermuk}
\hskip -3cm
  \vcenter{\xymatrix@!0@C=24mm@R=26mm{
    &&    & 
      \Big(X\times_B C_b,\, {\cal P}_B\otimes \pi_2^* {\cal N}_b \otimes \pi_1^* V_a\Big)
      \ar[dr]_{\pi_2}
      \ar[dl]^{\pi_1}
      \\
       && 
      \Big( X,\, V_a\otimes V_b \Big) & & 
      \Big( C_b, \, \pi_{2*} ({\cal P}_B\otimes \pi_1^* V_a) \otimes {\cal N}_b  \Big) \\
  }}
\end{equation}
Since the map $\pi_1$ is finite, i.e. its fiber consists  of $n_a$ points,
the Leray spectral sequence for $\pi_1$ reduces to the following relation for the cohomology classes
\bea
\label{lerayone}
     H^i(X\times_B C_b, {\cal P}_B\otimes \pi^\star_2 {\cal N}_b \otimes \pi_1^* V_a)=
        H^i(X,V_a\otimes V_b)\quad {\rm for}\ i=\{0,1,2,3\}.
\eea
We will now apply the Leray sequence to the projection $\pi_2 :X\times_B C_a \to C_a$. To compute $\pi_{2*}\left({\cal P}_B\otimes \pi_2^* {\cal N}_b \otimes \pi_1^* V_a\right)$ consider its restriction to a point $x \in C_b$
\begin{eqnarray}
\pi_{2*}\left({\cal P}_B\otimes \pi_2^* {\cal N}_b \otimes \pi_1^* V_a\right)|_{x} &= &H^0(\left({\cal P}_B\otimes \pi_2^* {\cal N}_b \otimes \pi_1^* V_a\right)|_{E_{\pi(x)\times x}}) \\ \nonumber
&= & H^0({\cal O}_{E_{\pi(x)}}(p-x)\otimes (\oplus_j {\cal O}_{E_{\pi(x)}}{(Q^a_i-p)}). \nonumber
\end{eqnarray}
Under the assumption of irreducibility of $C_a$ and $C_b$ this vanishes clearly for generic $x$. Therefore $\pi_{2*}\left({\cal P}_B\otimes \pi_2^* {\cal N}_b \otimes \pi_1^* V_a\right)$ vanishes identically, and we find
$$
     H^i(X\times_B C_a, {\cal P}_B\otimes \pi^\star_2 {\cal N}_b \otimes \pi_1^* V_a)=
        H^{i-1}(C_b,R^1\pi_{2*}\left({\cal P}_B\otimes \pi_2^* {\cal N}_b \otimes \pi_1^* V_a\right)),\; i=1,2.
$$
Combining this result with (\ref{lerayone}) gives
\bea
H^i(X,V_a\otimes V_b)=H^{i-1}(C_b,R^1\pi_{2*}\left({\cal P}_B\otimes \pi_2^* {\cal N}_b \otimes \pi_1^* V_a\right)),\; i=1,2.
\eea
Note that it follows also from considerations above that $R^1\pi_{2*}\left({\cal P}_B\otimes \pi_2^* {\cal N}_b \otimes \pi_1^* V_a\right)$
actually has support only on $\tau C_a \cap C_b$. Hence we define a rank one sheaf on $D$ 
$$
i_*{\cal L}=R^1\pi_{2*}\left({\cal P}_B\otimes \pi_2^* {\cal N}_b \otimes \pi_1^* V_a\right),
$$ 
where $i: D \to C_b$ denotes the inclusion map. Using Grothendieck-Riemann-Roch theorem for $\pi_2$ we can compute the Chern classes of $i_*\cal{L}$.
At zero order we find
\bea
             {\rm ch}_0({R^1\pi_{2*}\left({\cal P}_B\otimes \pi_2^* {\cal N}_b \otimes \pi_1^* V_a\right)})&=&\pi_{2\star}\left[ c_1(W) +
               {n_b\over 2} c_1(X\times_B C_a) \right] \\
                                         &=& \pi_{2\star}\left[ \Delta-\sigma_1\right ] =0. \nonumber
\eea 
This was expected, since $R^1\pi_{2*}\left({\cal P}_B\otimes \pi_2^* {\cal N}_b \otimes \pi_1^* V_a\right)$ is supported only on the curve $D$.
Similarly, at first order  we get
\bea
\label{chernone}
          c_1(i_*{\cal L})=[C_b\cdot C_a],
\eea
in agreement with $[\tau C_a]= [C_a]$ and $[D]=[C_b\cdot C_a]$.
For the second Chern class  we  obtain
\bea
\label{weisse}
{\rm ch}_2(i_*{\cal L})&=&-\lambda_a (-n_a\, \sigma_2+ \pi^* \eta_a-n_a\pi^*  c_1(B) )\cdot C_a\cdot C_b \nonumber \\
   &&      -\lambda_b (-n_b\, \sigma_2+\pi^* \eta_b-n_b\pi^*  c_1(B) )\cdot
   C_a\cdot C_b   \\
        && +\left( {1\over n_a} c_1(\zeta_a) + {1\over n_b} c_1(\zeta_b)
   \right) \cdot C_a\cdot C_b   \\
             &&+  {1\over 2}\, C_a\cdot C_a\cdot C_b - {pts}.\nonumber
\eea
Here the number of points $pts\not\subset
 C_a\cap C_b$ is  given by  $pts=(\sigma_2 +c_1(B))\cdot c_1(B)\cdot C_a=\eta_a\cdot c_1(B)\ge 0 $.

The interpretation of this result is as follows: 
First, the additional class of points
appearing in  (\ref{weisse}), as observed in \cite{Curio:1998vu},  reflects the fact that there exist 
point-like singularities in $X\times_B C_a$. This happens when the discriminant locus
meets the branch locus of $C\to B$. These can be blown up leading to changes in the 
Chern classes such that for the simplest choice of bundle resolution
this extra term disappears \cite{Curio:1998vu}.

We are left with computing the Chern classes of $\cal L$. Using Grothendieck-Riemann-Roch for the map $i: D \to C_b$, one finds that $\cal L$ has rank one and that
\bea
         c_1({\cal L}) = {\rm ch}_2(i_* {\cal L})+{1\over 2}\, C_a\cdot  
                     C_a\cdot C_b.
\eea
As can easily be verified, this implies that
\bea
 c_1({\cal L}) = c_1({\cal N}_{a}\otimes {\cal N}_{b} \otimes \pi^*K_B)
    \vert_{D} 
\eea
thus proving our claim.

\noindent
To summarize:
\vskip 0.3cm
\noindent
\hspace*{-9pt}
\fbox{
\begin{minipage}{\textwidth}
The non-vanishing cohomology classes of the tensor product of two  bundles
defined via the spectral cover method can
be computed from the cohomology classes of a certain line bundle on the
intersection curve of the two spectral surfaces:
\bea
    H^{i+1}(X,V_a\otimes V_b)=H^i(\tau C_a\cap C_b,\, \tau^*{\cal N}_{a}\otimes {\cal N}_{b} \otimes K_B),   \quad {\rm for}\ i=0,1.
\eea
\end{minipage}
}
\vskip 0.5cm
\noindent
Consistently, the direct Riemann-Roch-Hirzebruch theorem on the support curve yields the correct 
Euler characteristic of the bundle $V_a\otimes V_b$ on $X$,
\bea
   -\chi(X,V_a\otimes V_b)= c_1(\tau^*{\cal N}_{a}\otimes {\cal N}_{b}\otimes K_B)_{C_a\cap C_b}-{1\over 2}
               (C_a+C_b)\cdot C_a\cdot C_b.
\eea
The computation of $H^i(X,V_a\otimes V_b)$ is therefore reduced to the
computation 
of the cohomology of a line bundle over the curve $\tau C_a\cap C_b$ which is
the 
restriction of a line bundle defined on $X$. The standard procedure in such
situations 
is to invoke a series of Koszul sequences relating the cohomology of the restriction 
${\cal N}_{a}\otimes {\cal N}_{b} \otimes K_B$ to that of line bundles on $X$. 
The Koszul sequences which do the job for us are displayed in appendix
\ref{appKos}. 
Also, the cohomology of line bundles of $X$ is  easy to compute in view of 
section \ref{sec_H(L} and with the help of the results of appendix \ref{app_dPr}.

\subsection{Cohomology classes $H^i(X,\bigwedge^2 V)$ and
  $H^i(X,{\bf S}^2 V)$}

In this section we  compute the cohomology of $\bigwedge^2 V$ and
${\bf S}^2 V$
for the case that $V$ is a vector bundle of rank $r$ defined by the
spectral cover construction. Since our result differs
from the one in \cite{Donagi:2004ia}, we present our derivation in some detail. 

To begin with, recall that generally
\bea\label{rwV}
         V\otimes V = \left[{\textstyle \bigwedge^2 V} \right]\oplus \left[{\textstyle {\bf S}^2 V}\right].
\eea
Using the results of the previous section for $V=V_a=V_b$, we can immediately
conclude
\bea
\label{homolhwedge}
     H^{i+1}(X,V\otimes V)=H^i(\tau C\cap C,\, \tau^*{\cal N} \otimes {\cal N}\otimes K_B),
\quad {\rm for}\ i=0,1.   
\eea
In the sequel we will again assume that indeed $\tau^*{\cal N} =  {\cal N}$. The zero and third order cohomology groups vanish and the righthand side of 
(\ref{homolhwedge}) can be computed using the Koszul sequences in appendix \ref{appKos}.

To proceed, we will again use 
the F-theory respectively Type IIB orientifold
intuition. In the orientifold limit of the dual F-theory, the orientifold
projection is simply $\Omega I_2 (-1)^{F_L}$, where $I_2$ denotes the holomorphic involution of 
the  fiber $T^2$. For a stack of $N$ D7-branes wrapping a four-cycle $C$ and carrying a $U(N)$ gauge group, 
matter fields transforming in the symmetric and anti-symmetric
representations of the gauge group are localized on the intersection of $C$ with the image of $C$ under the holomorphic involution $I_2$.
On the heterotic side the orientifold projection maps precisely to the involution $\tau$ discussed at the end of  section \ref{seccy}.
Therefore the matter is localized on the curve
\bea
D=\tau C \cap C.
\eea
Note that $D$ is invariant under $\tau$. To study the curve $D$, we 
consider the fiberwise decomposition of (\ref{rwV}) for a generic fiber $f_b$
\begin{eqnarray}
\label{tensorprod}
\left( \oplus_i {\cal O}_{f_b}(Q_i-p)\right)  \otimes \left( \oplus_j {\cal O}_{f_b}(Q_j-p) \right)&=&\left( \oplus_{i<j}{\cal O}_{f_b}(Q_i+Q_j-2p)\right) \\
&\oplus& \left(\oplus_{i\leqslant j}{\cal O}_{f_b}(Q_i+Q_j-2p)\right). \nonumber
\end{eqnarray}
We recall from the previous section that the condition on the eigenvalues  $Q_i$ of $V$ on a fiber $f_b$ to be in $D$ are
\bea
Q_j=-Q_i.
\eea
Let us assume $i=j$. Then we find $2Q_i=0$. This is the intersection of $C$ with the zero section $\sigma$ and the intersection of $C$ with the triple section $\sigma_t$ describing points of order two on elliptic fiber. Let us assume that $i\neq j$. Then $Q_i=-Q_j$ implies $Q_j=-Q_i$, hence this fiber contains two  points of $D$. We conclude  that $D$ generically consists of three components
\bea
D=D \cap \sigma + D \cap \sigma_t +D'=C \cap \sigma + C \cap \sigma_t +D'.
\eea
However, the fixed point locus  $C\cap \sigma_{\tau}= C \cap \sigma + C \cap
\sigma_t$ intersects $D'$ in 
\bea
\label{ramif}
R=(C-\sigma_\tau)\cdot C\cdot \sigma_\tau
\eea
points.
It follows that for the line bundle $L^2={\cal N}^2\otimes K_B$ we have the 
exact sequence
\bea
    0\to L^2\otimes {\cal O}(-R)\vert_{C\cap \sigma_{\tau}}\to L^2\vert_{\tau C\cap
      C} \to L^2\vert_{D'}\to 0 
\eea
implying the corresponding long exact sequence in cohomology.
We now have to
split each appearing cohomology group  into its $\tau$ symmetric and anti-symmetric
component.
Clearly, the fixed point locus $C\cap \sigma_{\tau}$
contributes entirely to the cohomology of ${\bf S}^2 V$ since we have just identified it with the points $i=j$ appearing in (\ref{tensorprod}). Thus
\bea
H^i_+(C\cap \sigma_\tau, L^2)=H^i(C\cap \sigma_\tau, L^2),\quad 
H^i_-(C\cap \sigma_\tau, L^2)=0.
\eea
This is consistent with the orientifold dual, where the fixed point locus of the involution
only contributes to anti-symmetric matter and therefore
to $H^i(X,{\bf S}^2 V)$. 
Therefore, we can conclude that
\bea
     H^{i+1}(X,{\textstyle \bigwedge^2 V})=H^i_- (D',L^2)
\eea
and that $H^{i+1}(X,{\bf S}^2 V)$ must be determined from the
exact sequence
\begin{equation}
  \label{llseq}
  \vcenter{\xymatrix@R=10pt@M=4pt@H+=22pt{
      0 \ar[r] & H^0(C\cap \sigma_{\tau}, L^2 \otimes {\cal O}(-R)) \ar[r] &
       H_+^0(\tau C\cap C, L^2) \ar[r] &
      H^0_+ (D',L^2)  
      \ar`[rd]^<>(0.5){}`[l]`[dlll]`[d][dll] & 
      \\
      & H^1(C\cap \sigma_{\tau}, L^2 \otimes {\cal O}(-R)) \ar[r] &
       H_+^1(\tau C\cap C, L^2)   \ar[r] &
      H^1_+ (D',L^2) \ar[r] &
      0
      \,
    }}
\end{equation}
What remains is to determine $H^i_\pm (D',L^2)$.

To obtain a numerical tool for  the computation of this splitting 
recall from the previous section that 
\bea
H^i(X,{\textstyle \bigwedge^2} V)=H^{i-1}(c_{\wedge^2 V}, {\cal N}_{\wedge^2 V}\otimes K_B|_{c_{\wedge^2 V}}),\;i=1,2,
\eea
where $c_{\wedge^2 V}$ denotes the intersection of the spectral cover of $\bigwedge^2 V$ with $\sigma$ and ${\cal N}_{\wedge^2 V}$ its spectral rank one sheaf. It is important to realize that the surjective map $\pi_{D'}: D' \to c_{\wedge^2 V}$, which is generically two-to-one, factors through 
\bea
\xymatrix{
D' \ar[r]^{\pi_{D'}} \ar[d]^m& c_{\wedge^2 V} \\
D'/\tau \ar[ru]^n
}
\eea
Here $D'/\tau$
is the normalization of $c_{\wedge^2 V}$. 
The canonical bundle of $D'$ has degree $C(C-\sigma_\tau)(2C -\sigma_\tau)$ and is related to the canonical bundle of $D'/\tau$  by
\bea
m^* K_{D'/\tau}= K_{D'}\otimes {\cal O}_{D'}(-R),
\eea
where $R$ is the ramification divisor (\ref{ramif}).
In particular, 
\bea
c_1(K_{D'/\tau})\vert_{D'/\tau}=\frac{1}{2}(C(C-\sigma_\tau)(2C-\sigma_\tau)-R).
\eea

Applying $m_*$ to ${L}^2$, we obtain a rank two vector bundle on 
$D'/\tau$ which splits into a sum of line bundles
\bea
{m_*} {L}^2 = L_i \oplus L_{a}.
\eea
The sections of $L_i$  and $L_a$ are invariant and anti-invariant under $\tau$, respectively. In particular,
\bea
m^* L_i = {L}^2
\eea
and 
\bea
H^i(D',L^2) = H^i(D',m^* L_i).
\eea
Therefore, in oder to compute the anti-invariant part of the cohomology of
${L}^2$ we have to compute $H^i(D'/\tau,L_{a})$. 
Clearly, at generic points $n_* L_a = {\cal N}_{{c}_{\wedge^2 V}}\otimes K_B$, which we expect to hold everywhere.
Using GRR for the map $m$, we find that 
\bea
c_1(L_a)\vert_{D'/\tau}=c_1\left({\cal N}\otimes K^{1/2}_B\right)\bigr\vert_{D'}+R/2.
\eea
Note that, since $\pi^*K_B|_{D'}$ is invariant under $\tau$, 
it is the pull-back of a bundle $K_B^{1/2}$ on $D'/\tau$. 

Let us  summarize our final formulas for  the cohomology groups
of the anti-symmetric and symmetric product bundles.
\vskip 0.5cm
\noindent
$\bullet$\ $H^i(X,\bigwedge^2 V)$
\vskip 0.3cm
\noindent
\hspace*{-9pt}
\fbox{
\begin{minipage}{\textwidth}
The non-vanishing cohomology groups of the bundle $\bigwedge^2 V$
can be computed from the cohomology groups of a line bundle 
$L_a$ on the quotient $D'/\tau$ by
\bea
     H^{i+1}(X,{\textstyle \bigwedge^2 V})=H^i_- (D',L^2)=H^i(D'/\tau,L_a), \quad
    {\rm for}\ i=0,1. \nonumber
\eea
with the first Chern class of $L_a$ given by
\bea
c_1(L_a)\vert_{D'/\tau}=c_1\left({\cal N}\otimes K^{1/2}_B\right)\bigr\vert_{D'}+R/2. 
\eea
\end{minipage}
}
\vskip 0.3cm
\noindent
Applying the RRH theorem  to this line bundle we find
\bea
   -\chi(X,{\textstyle \bigwedge^2 V})= 
   c_1\left({\cal N}\otimes K_B^{1/2}\right)\bigr\vert_{D'} -{1\over 2}\,
               C \cdot C\cdot (C - \sigma_\tau ),
\eea
an important consistency check of our computation.
\vskip 0.5cm
\noindent
$\bullet$\ $H^i(X,{\bf S}^2 V)$
\vskip 0.3cm
\noindent
\hspace*{-9pt}
\fbox{
\begin{minipage}{\textwidth}
The non-vanishing cohomology groups of the bundle ${\bf S}^2 V$,
\bea
H^{i+1}(X,{\bf S}^2 V) = H^i_+(\tau C \cap C,L^2),
\eea
can be computed from the sequence (\ref{llseq}) with 
\bea
     H^i_+ (D',L^2)=H^i(D'/\tau,L_i), \quad
    {\rm for}\ i=0,1, \nonumber
\eea
and  the first Chern class of $L_i$ is given by
\bea
c_1(L_i)\vert_{D'/\tau}=c_1\left( {\cal N}\otimes K^{1/2}_B\right)\bigr\vert_{D'}.
\eea
\end{minipage}
}
\vskip 0.3cm
\noindent
In particular, this implies together with the sequence (\ref{llseq}) that $-\chi(X,{\textstyle {\bf S}^2 V}) = \chi(C\cap \sigma_{\tau}, L^2 \otimes {\cal O}(-R)) + \chi_+ (D',L^2)$, and again the RRH theorem  consistently gives
\bea
   -\chi(X,{\textstyle {\bf S}^2 V})= 
   c_1\left({\cal N}\otimes K_B^{1/2}\right)\bigr\vert_{\tau C\cap C} -{1\over 2}\,
               C \cdot C\cdot (C + \sigma_\tau ).
\eea

\section{Heterotic flipped $SU(5)$ GUT models}
\label{sec_GUT}

Having presented the mathematical framework for the computation of the complete massless
spectrum of heterotic string compactification with $U(N)$ bundles, we are now in a position to apply these techniques to heterotic model building.
After briefly summarizing the way flipped $SU(5)$ vacua were obtained in \cite{Blumenhagen:2006ux}
we present a new fully consistent three-generation string vacuum. 
For this phenomenologically promising model, we exemplify the methods
developed in the first part of this paper and compute the complete massless spectrum.

\subsection{$SU(4)\times U(1)$ bundles}

We consider a bundle with structure group $SU(4)\times U(1)$
on a Calabi-Yau $X$ including cases with $\pi_1(X)=0$. 
Such types of construction were considered in \cite{Distler:1987ee}
before and further details of this particular one can be found in \cite{Blumenhagen:2005ga,Blumenhagen:2006ux}.\footnote{For different aspects of this and related constructions see \cite{Tatar:2006dc}. Recent investigations of heterotic K3 compactifications with line bundles are performed in \cite{Honecker:2006dt,Gabineu}. Some previous results on heterotic $U(N)$ bundles in six and five-dimensional compactifications appear in \cite{Green:1984bx,Aldazabal:1996du} and \cite{Lukas:1999nh}, respectively.}

More precisely, our starting point is the direct sum 
\bea    \label{BundleU4a}
    W_1=V_1\oplus L^{-1} \quad {\rm with}\ c_1(V_1)=c_1(L), \, \, \,  \mbox{rank}(V)=4, 
\eea
where $V_1$ is $U(4)$ vector bundle, $L$ is a complex line bundle and the structure group of $W_1$ is $ G_1 = SU(4)\times U(1)$ due to the constraint $c_1(V_1)=c_1(L)$.
$G_1$ can now
be embedded into an $SU(5)$ subgroup of the first $E_8$ such that its commutant in $E_8$
is  $SU(5)\times U(1)_1$. For the details of embeddings of this type we refer to \cite{Blumenhagen:2005ga,Thesis}. The decomposition of the adjoint ${\bf 248}$ of $E_8$, 
\bea
\label{breakingSU5}
{\bf 248} 
 \stackrel{SU(4) \times SU(5) \times U(1)_{1}}{\longrightarrow}
\left\{\begin{array}{c}
({\bf 15},{\bf 1})_0 \\
({\bf 1}, {\bf 1})_0 + ({\bf 1},{\bf 10})_{-4} + ({\bf 1}, \ov{\bf 10})_{4} + ({\bf 1}, {\bf 24})_0 \\
(\4,{\bf 1})_{5} + (\4, \ov{\5})_{-3} + (\4,{\bf 10})_{1} \\
(\ov{\4},{\bf 1})_{-5} + (\ov{\4}, \5)_{3} + (\ov{\4},\ov{\bf 10})_{-1} \\
(\6,\5)_{-2} + (\6,\ov{\5})_{2}
\end{array}\right\}, 
\eea
reveals that the spectrum is precisely that of flipped $SU(5) \times U(1)_X$ \cite{Barr:1981qv} provided we guarantee that the abelian gauge group remains massless in the process of Green-Schwarz type anomaly cancellation. 

In \cite{Blumenhagen:2006ux} it was proposed to embed a second line bundle
into the other $E_8$ such that a linear combination of the two observable
$U(1)$'s remains massless.
It turns out, however, that in order to construct models with precisely the Standard Model matter content and no further non-chiral matter, it is more convenient to invoke in the hidden $E_8$ a slightly more sophisticated structure than the one detailed in \cite{Blumenhagen:2006ux}. Namely, we can consider the second simplest embedding
\bea
U(2) \times U(1) \subset E_8 \rightarrow E_6 \times U(1)_2
\eea
inducing the decomposition

\bea
\label{breaking}
{\bf 248} 
 \stackrel{SU(2) \times E_6 \times U(1)_2}{\longrightarrow}
\left\{\begin{array}{c}
({\bf 1}, {\bf 78})_0  \\
({\bf 1}, {\bf 1})_0 + (\3,{\bf 1})_0 + ({\bf 2}, {\bf 1})_{3} + ({\bf 2},{\bf 1})_{-3}  \\
({\bf 2},{\bf 27})_{1} +({\bf 1},{\bf 27})_{-2}  \\
({\bf 2},{\bf \overline {27 }})_{-1} +({\bf 1},{\bf \overline {27}})_{2} \\
\end{array}\right\}. 
\eea

The bundle we embed into the second $E_8$ is of the form
\bea
W_2 = V_2 \oplus L^{-1},   \quad \quad c_1(V_2) = c_1(L),
\eea
where we stress that the line bundle $L$ is the same as the one appearing in (\ref{BundleU4a}). The resulting chiral spectrum in the second $E_8$ is counted by the cohomology groups listed
in table \ref{cohom_hidden2}.
\begin{table}[htb]
\renewcommand{\arraystretch}{1.5}
\begin{center}
\begin{tabular}{|c|c|}
\hline
\hline
$E_6\times U(1)_2$ & cohomology  \\
\hline \hline
${\bf 1}_{3}$ & $V_2 \otimes L $  \\
${\bf 27}_{1}$ & $V_2$   \\
${\bf 27}_{-2}$ & $  L^{-1} $ \\
\hline
\end{tabular}
\caption{\small Massless spectrum of $H=E_6\times U(1)_2$ models.}
\label{cohom_hidden2}
\end{center}
\end{table}

The unitary vector bundles $V_1$, $V_2$ and the complex line bundle $L$ are subject to a number of constraints to guarantee that the model constitutes a well-defined string vacuum with the desired spectrum.
The non-trivial Bianchi identity for the three-form field strength ensuring anomaly cancellation translates into the following cohomological constraint on the second Chern classes of the vector bundles and the tangent bundle of the Calabi-Yau,
\bea
\label{TADSU(3)}
{ \rm ch}_2(V_1) +  { \rm ch}_2(V_2) + c_1^2(L) - \sum_a N_a {\overline \gamma}_a = -c_2(X).
\eea
Here we allowed for the presence of stacks of $N_a$ space-time-filling five-branes wrapping the holomorphic two-cycles $\Gamma_a$ dual to the four-form ${\overline \gamma}_a$ on $X$.

According to the reasoning detailed in \cite{Blumenhagen:2006ux}, the combination 
\bea
U(1)_f = -\frac{1}{2}\left(U(1)_1 - \frac{5}{3}U(1)_2 \right)
\eea
remains massless if the following conditions are satisfied
\bea
\label{masslesscond}
 &&\int_X  c_1(L)\wedge c_2(V_1) =0, \quad\quad  \int_X  c_1(L)\wedge c_2(V_2) =0, 
\nonumber \\
&&\int_{\Gamma_a} c_1(L)=0 \quad {\rm for \, \,  all\, \,  M5\, \, branes}.
\eea
In this case, the one-loop correction (\ref{Lstable}) to the slope vanishes and we are in the fortunate situation of (\ref{l=m}), i.e. that $\mu$-stability of $V_1$ and $V_2$, together with the DUY condition (\ref{DUY}), guarantees supersymmetry in the perturbative regime.

Chiral right-handed electrons  from the second $E_8$ with non-MSSM Yukawa couplings are absent if in addition
\bea
\label{rhelectrons}
          \int_X c^3_1(L)=0.
\eea
In this case one obtains no chiral matter at all resulting from the hidden $E_8$. 
The complete spectrum can be found in table \ref{spec_E6}.
\begin{table}[htb]
\renewcommand{\arraystretch}{1.5}
\begin{center}
\begin{tabular}{|c||c|c|c|}
\hline
\hline
$SU(5)\times U(1)_X \times E_6$ & Cohomology & $\chi$  & SM part. \\
\hline \hline
$ ({\bf 10},{\bf 1})_{1\over 2}$ & $H^i(V_1)$ & $g$ & $(q_L,d^c_R,\nu^c_R)+[H_{10}+\overline H_{10}]$ \\
$( {\bf 10},{\bf 1})_{-2}$ & $H^i(L^{-1})$ & $0$ &$-$ \\
\hline
$(\overline \5,{\bf 1})_{-{3\over 2}}$ & $H^i(V_1\otimes L^{-1})$ & $g$ & $(u^c_R,l_L)$ \\
$(\overline \5,{\bf 1})_{1}$ & $H^i(\bigwedge^2 V_1)$ & $0$ & $[(h_3, h_2)+(\overline h_3, \overline h_2)]$ \\
\hline

 $({\bf 1},{\bf 1})_{{5\over 2}}$ & $H^i(V_1\otimes L)+H^i(V_2^\vee \otimes L^{-1})$ & $g$ & $e^c_R$ \\
\hline
$({\bf 1},{\bf 27})_{{5\over 6}}$ & $H^i(V_2)$ & $0$ & $-$ \\
$({\bf 1},{\bf 27})_{-{5\over 3}}$ & $H^i(L^{-1})$ & $0$ & $-$ \\
\hline
\end{tabular}
\caption{\small Massless spectrum of $H=SU(5)\times U(1)_X$ models with hidden $E_6$ symmetry;
      $g={1\over 2}\int_X c_3(V_1)$.}
\label{spec_E6}
\end{center}
\end{table}

Whereas the net number of chiral generations is simply given by ${1\over 2}\int_X c_3(V_1)$ once the constraints (\ref{TADSU(3)}), (\ref{masslesscond}), (\ref{rhelectrons}) are satisfied, the vector-like matter is described by the cohomology groups listed in table \ref{spec_E6}.
Our task is therefore to find  
stable vector $U(4)$ and $U(2)$ bundles $V_1$ and $V_2$ as well as a line bundle $L$  subject to the constraints discussed such that
\begin{itemize}
\item $H^*(X,V_1)=(0,1,4,0)$ for precisley one pair of GUT Higgs and 3 generations of $(q_L,d^c_R,\nu^c_R)$,
\item  $H^*(X,V_1\otimes L^{-1}) = (0,0,3,0)$  for 3 generations of $(u^c_R,l_L)$, 
\item  $H^*(X,V_1\otimes L) = (0,0,3,0)$  for 3 generations of  $e^c_R$ and 
\item  $H^*(X,L^{-1})=H^*(X,V_2) = (0,0,0,0) = H^*(X,V_2 \otimes L)$ in order to avoid all kinds of non-chiral exotic matter.
\end{itemize}
The number of Higgs pairs is then determined by $H^*(X,\bigwedge^2 V_1)$.

\subsection{A three-generation model from extensions}

We now provide an example of the flipped $SU(5)$ framework whose spectrum comes remarkably close to the Standard Model.
Our Calabi-Yau manifold $X$ is elliptically fibered over the basis $B$, which we take to be d$P_4$.

Let us start with the visible $E_8$, into which we embed the direct sum $W_1= V_1 \oplus L^{-1}$ of a stable $U(4)$ bundle $V_1$ and the line bundle $L$ with $c_1(V_1) = c_1(L)$.
For $L$ we choose the pull-back of a line bundle on dP$_4$ with first Chern class 
\bea
        c_1(L)= -E_1 + E_4.
\eea 
Since L is the pull-back of a line bundle on the base space, clearly
\bea
        \int_X c_1(L)^3 = 0
\eea
and we therefore have no further contributions to the right-handed electrons.

$V_1$ is constructed as the extension of two stable $U(2)$ bundles $V_a$ and $V_b$,
\bea
\label{ext_2}
        0 \rightarrow V_a \rightarrow V_1 \rightarrow V_b \rightarrow 0,
\eea
where $V_a$ and $V_b$ are obtained via the spectral cover
construction. Concretely, 
the defining data is in the notation of section 2
\begin{equation}
\begin{array}{lll}
   \lambda_a = 0, &c_1(\zeta)_a = l - 2 E_2,    &           \eta_a = 12l - 5E_1 -5E_2 -3E_3 - 5E_4, \nonumber \\
   \lambda_b = 0, &c_1(\zeta)_b = -l - E_1 + 2 E_2 + E_4, & \eta_b = 10l - 4E_1 - E_2 - 3E_3 - 4E_4. \nonumber
\end{array}
\end{equation}
Clearly $c_1(V_1) = c_1(V_a) + c_1(V_b) = - E_1 + E_4 = c_1(L)$, as required.

In the hidden $E_8$ we embed $W_2 = V_2 \oplus L^{-1}$, where the $U(2)$ bundle $V_2$ is given by the data
\bea
       \lambda_2 = 0, \quad  c_1(\zeta)_2 = -E_1 + E_4,  \quad \eta_2=7l -2E_1 -3 E_2 -3 E_3 - 2 E_4, 
\eea
again satisfying $c_1(V_2) = c_1(L)$.
One can check that each of the bundles $V_a$, $V_b$ and $V_2$ are stable in that they satisfy the corresponding criteria described in section \ref{secscc}.

\vspace{10pt}

\underline{Proof of stability} 
\vspace{10pt}

We now prove stability of the bundle $V_1$ defined via the extension (\ref{ext_2}). As discussed in section 2.5 and in appendix A, this amounts to showing that the extension is non-split and that $\mu(V_a) < \mu(V_1) = 0$ for a K\"ahler class inside the K\"ahler cone.
Starting with this latter constraint, we parameterise the K\"ahler form $J$ on $X$ as $J= \ell_s^2(r_{\sigma} \, \sigma + \pi^*(r_0 l + \sum_{i=1}^4 r_i E_i))$. Note that $r_{\sigma}$ measures the area of the fibre.
The values for $r_{\sigma}$, $r_0$ and $r_i$ have to be such that $J$ lies inside the K\"ahler cone. For the numerical constraints following from this requirement we refer e.g. to appendix A of \cite{Blumenhagen:2005zg}. One may check that they are satisfied  for the choice
\bea
\label{Kahlerpara}
        0 < r_\sigma < 4\rho,\quad r_0=6\rho,\quad r_1=-2\rho,\quad r_2=-3\rho,\quad r_3=-2\rho,\quad r_4=-2\rho,
\eea
where $\rho > 0$. Note that for this choice, the DUY-condition (\ref{DUY}) for $V_1$ and $V_2$ is fulfilled and therefore $\mu(V_{1,2}) = 0$. Stability of each of the bundles $V_a$, $V_b$ and $V_2$ requires furthermore that $r_{\sigma} < \epsilon$ for some (in general unknown) critical value of $\epsilon$. With the help of the intersection form on the basis we readily compute that
\bea
        \int_X J \wedge J \wedge c_1(V_a) &=& \ell_s^4 \left( -r_{\sigma}^2 + 2r_{\sigma}( r_0 + 2r_2)\right).
\eea

Restricting oneself for simplicity to the parameter space in (\ref{Kahlerpara}) one concludes that 
$\mu(V_a) < 0$ translates into
\bea
        r_{\sigma}^2 > 0,
\eea
which is always true.
Most importantly, stability of $V_1$ does therefore not set a lower bound on $r_{\sigma}$ so that we can indeed take it smaller than any critical $\epsilon$ required by the spectral cover construction.

The second part of the stability condition on $V_1$ requires the computation of $H^1(X, V_a \otimes V_b^\vee )$.
According to equation (\ref{homolh}), $H^i(X, V_a \otimes V_b^\vee) = H^{i-1}(C_a \cap C_b, {\cal L}|_{C_a \cap C_b})$ for $i=1,2$, where ${\cal L} = {\cal N}_{V_a} \otimes {\cal N}_{V_b^\vee} \otimes K_B$. As discussed in appendix \ref{appKos}, we have to invoke a series of three Koszul sequences in which four line bundles on $X$ appear:
$\cL, \cL \otimes {\cal O}(-\cC_a),\cL \otimes {\cal O}(-\cC_b), \cL \otimes {\cal O}(-\cC_a-\cC_b)$.
Their cohomology groups are easily determined once we know their first Chern classes. From the definition of the spectral line bundle (\ref{linebundle}) and the concrete bundle data we find
\bea
        c_1({\cal N}_{V_a}) &=& \sigma + \pi_C^*\left( \frac{1}{2}(\eta_a + c_1(B)) + \frac{1}{n}c_1(\zeta)_a\right), \nonumber \\
        c_1({\cal N}_{V_b^\vee}) &=& \sigma + \pi_C^*\left( \frac{1}{2}(\eta_b + c_1(B)) - \frac{1}{n}c_1(\zeta)_b\right).
\eea
Thus,
\bea
        c_1(\cL) &=& 2 \sigma + \pi_C^*(12l - 4E_1 - 5E_2 -3E_3 -5E_4),  \nonumber \\
        c_1(\cL \otimes {\cal O}(-\cC_a)) &=& \pi_C^*(E_1), \nonumber \\
        c_1(\cL \otimes {\cal O}(-\cC_b)) &=& \pi_C^*(2l - 4E_2 - E_4), \nonumber \\
        c_1(\cL \otimes {\cal O}(-\cC_a-\cC_b)) &=& -2\sigma + \pi_C^*(-10l + 5E_1 + E_2 + 3E_3 + 4E_4)
\eea
with Hodge numbers
\bea
        H^*(X,\cL) &=& (57,0,0,0), \nonumber \\ 
        H^*(X,\cL \otimes {\cal O}(-\cC_a)) &=& (1,0,0,0), \nonumber \\ 
        H^*(X,\cL \otimes {\cal O}(-\cC_b)) &=& (0,5,6,0), \nonumber \\
        H^*(X,\cL \otimes {\cal O}(-\cC_a-\cC_b)) &=& (0,0,0,39).
\eea
In all we find 
\bea
        H^*(X, V_a \otimes V_b^\vee) = (0, 61 , 45, 0)
\eea
and therefore the extension is non-split. This completes the proof of stability of $V_1$.

\vspace{10pt}

\underline{Checking the consistency conditions} 
\vspace{10pt}

In section \ref{secscc}, we listed the Chern characters for spectral cover bundles (see (\ref{Chern1})). The result for the two vector bundles $V_a$ and $V_b$ in this example is
\bea
        \ch_1(V_a) &=& l - 2 E_2,\nonumber\\
        \ch_2(V_a) &=& \sigma\pi^*(-12l + 5E_1 + 5E_2 + 3E_3 + 5E_4) + \frac{13}{2}F,\nonumber\\
        \ch_3(V_a) &=& -1,
\eea
\bea
        \ch_1(V_b) &=& -l - E_1 + 2 E_2 + E_4,\nonumber\\
        \ch_2(V_b) &=& \sigma\pi^*(-10l + 4E_1 + 1E_2 + 3E_3 + 4E_4) + \frac{11}{2}F,\nonumber\\
        \ch_3(V_b) &=& 4.
\eea
$V_1$ being the extension of $V_b$ by $V_a$ its total Chern character is the sum of the total Chern characters of $V_a$ and $V_b$. Thus,
\bea
        \ch_1(V_1) &=& - E_1 + E_4,\nonumber\\
        \ch_2(V_1) &=& \sigma\pi^*(-22l + 9E_1 + 6E_2 + 6E_3 + 9E_4) + 12F,\nonumber\\
        \ch_3(V_1) &=& 3.
\eea
The Chern classes are then
\bea
        c_1(V_1) &=& - E_1 + E_4,\nonumber\\
        c_2(V_1) &=& \sigma\pi^*(22l - 9E_1 - 6E_2 - 6E_3 - 9E_4) - 13F,\nonumber\\
        c_3(V_1) &=& 6.
\eea
From the second Chern class, one can easily read off that the first line of the masslessness conditions (\ref{masslesscond}) holds.

For the $U(2)$-bundle in the hidden $E_8$, $V_2$, the Chern characters come out to be
\bea
        \ch_1(V_2) &=& - E_1 + E_4,\nonumber\\
        \ch_2(V_2) &=& \sigma\pi^*(-7l + 2E_1 + 3E_2 + 3E_3 + 2E_4) + F,\nonumber\\
        \ch_3(V_2) &=& 0.
\eea
To satisfy the tadpole condition (\ref{TADSU(3)}), the Poincar\'e dual four-form of the two-cycles, the five-branes are wrapping must be:
\bea
        \sum_a N_a \overline{\gamma_a} &=& \ch_2(V_1) + \ch_2(V_2) + c_1(L)^2 + c_2(X) \nonumber\\
        &=& \sigma\pi^*(7l - E_1 - 3E_2 - 3E_3 - E_4) + 73F.
\eea
This can be decomposed in a sum of positive multiples of irreducible cycles, for example:
\begin{table}[h]
\renewcommand{\arraystretch}{1.5}
\begin{center}
\begin{tabular}{|c||c|c|}
\hline
\hline
$a$ & $N_a$ & $\overline{\gamma_a}$\\
\hline \hline
$1$ & $1$ & $\sigma\pi^*(l-E_1-E_4)$\\
\hline
$2$ & $6$ & $\sigma\pi^*(l-E_2-E_3)$\\
\hline
$3$ & $3$ & $\sigma\pi^*(E_2)$\\
\hline
$4$ & $3$ & $\sigma\pi^*(E_3)$\\
\hline
$5$ & $73$ & $F$\\
\hline
\hline
\end{tabular}
\end{center}
\end{table}

With this decomposition, it is easy to see that the second line in equation (\ref{masslesscond})
\bea
        \int_{\Gamma_a} c_1(L) = \int_X c_1(L)\wedge\overline{\gamma_a} = 0
\eea
indeed holds for all components.

\vspace{10pt}

\underline{Computation of the massless spectrum} 
\vspace{10pt}

As mentioned in section \ref{secext}, we can calculate the cohomology groups $H^*(X,V_1)$ by the long exact sequence in cohomology, induced by (\ref{ext_2}):
\bea
0&\rightarrow& H^0(X,V_a) \rightarrow H^0(X,V_1) \rightarrow H^0(X,V_b) \rightarrow \nonumber \\
&\rightarrow& H^1(X,V_a) \rightarrow H^1(X,V_1) \rightarrow H^1(X,V_b) \rightarrow \ldots
\eea
For the cohomology groups $H^*(X,V_a)$ and $H^*(X,V_b)$, one can use again the method described in section \ref{sec_VaVb} by considering the tensor product with the trivial vector bundle ${\cal O}_X$ respectively. The results are
\bea
        H^*(X,V_a) &=& H^*(X,V_a\otimes {\cal O}_X) = (0,1,1,0), \nonumber\\
        H^*(X,V_b) &=& H^*(X,V_b\otimes {\cal O}_X) = (0,0,3,0)
\eea
and therefore
\bea
        H^*(X,V_1) = (0,1,4,0).
\eea
The exact sequence from the extension (\ref{ext_2}) remains exact upon tensoring every element with a line bundle. Thus, we can calculate the cohomology groups $H^*(X,V_1\otimes L)$ and $H^*(X,V_1 \otimes L^{-1})$ by the long exact sequence in cohomology, induced by the tensored short exact sequence. We find
\bea
        H^*(X,V_a\otimes L) &=& (0,0,0,0), \nonumber\\
        H^*(X,V_a\otimes L^{-1}) &=& (0,0,0,0), \nonumber\\
        H^*(X,V_b\otimes L) &=& (0,0,3,0),\nonumber\\
        H^*(X,V_b\otimes L^{-1}) &=& (0,0,3,0), \nonumber\\
\eea
yielding
\bea
        H^*(X,V_1\otimes L) &=& (0,0,3,0), \nonumber\\
        H^*(X,V_1\otimes L^{-1}) &=& (0,0,3,0).
\eea
For the computation of the cohomology groups $H^*(X,V_1)$, we use that the short exact sequence (\ref{ext_gen}) induces the following set of exact sequences
\bea
\label{big_seq}
              \vcenter{
                \hbox{$\phantom{0\to \bigwedge^2 V_a \to i} 0
                       \phantom{aaai\to } 0
                      \phantom{\to 0}$}
                 \hbox{$\phantom{0\to \bigwedge^2 V_a \to ,} \downarrow
                      \phantom{aai\to } \downarrow
                      \phantom{\to 0}$}
                \hbox{$0\to \bigwedge^2 V_a \to Q_1\to V_a\otimes V_b\to 0$}
                \hbox{$\phantom{0\to \bigwedge^2 V_a \to ,} \downarrow
                      \phantom{aai\to } \downarrow
                      \phantom{\to 0}$}
                \hbox{$0\to \bigwedge^2 V_a \to \bigwedge^2 V_1 \to Q_2\to 0$}
                \hbox{$\phantom{0\to \bigwedge^2 V_a \to ,} \downarrow
                      \phantom{aai\to } \downarrow
                      \phantom{\to 0}$}
                \hbox{$\phantom{0\to \bigwedge^2 V_a \to } \bigwedge^2 V_b
                      \phantom{\to } \bigwedge^2 V_b
                      \phantom{\to 0}$}
                \hbox{$\phantom{0\to \bigwedge^2 V_a \to ,} \downarrow
                      \phantom{aai\to } \downarrow
                      \phantom{\to 0}$}
                 \hbox{$\phantom{0\to \bigwedge^2 V_a \to i} 0
                       \phantom{aaai\to } 0
                      \phantom{\to 0}$}}
\eea
Since $V_a$ and $V_b$ are bundles of rank $2$, their anti-symmetric product is actually a line bundle and its cohomology can be computed using the method described in section \ref{sec_H(L}. The result is
\bea
\label{cohom_wedge}
        H^*(X,{\textstyle \bigwedge}^2 V_a) &=& (0,0,1,0), \nonumber\\
        H^*(X,{\textstyle \bigwedge}^2 V_b) &=& (0,2,1,0), \\
        H^*(X,V_a \otimes V_b) &=& (0,53-\rank f,53-\rank f,0), \nonumber\\
\eea
where $f$ is the map $f:H^1(X,{\cal L}\otimes{\cal O}(-{\cal C}_a)|_{{\cal C}_b}) \rightarrow H^1(X,{\cal L}|_{{\cal C}_b})$ appearing in the Koszul sequences. These spaces are both one-dimensional, so $f$ might in principle have rank $0$ or $1$. Resolving the various induced long exact sequences in cohomology in (\ref{big_seq}) gives
\bea
        H^*(X,Q_1) &=& (0,53-\rank f - \rank g, 54 - \rank f - \rank g, 0),  \nonumber\\
        H^*(X,Q_2) &=& (0,55-\rank f - \rank h, 54 - \rank f - \rank h, 0), \nonumber\\
\eea
where $g$ and $h$ are the maps $g:\, H^1(X, V_a\otimes V_b) \rightarrow H^2(X, \wedge^2 V_a)$, $h:\, H^1(X, \wedge^2 V_b) \rightarrow H^2(X, V_a \otimes V_b)$. From the dimensions of their image and domain, (\ref{cohom_wedge}), one can read off that their ranks can at most lie in the ranges $[0,1]$ and $[0,2]$ respectively.
Using these results, the exact sequence for $\bigwedge^2 V_1$ gives
\bea
        H^*(X,{\textstyle \bigwedge}^2 V_1) = (0, 55 - \rank f - \rank g - \rank i, 55 - \rank f - \rank g - \rank i, 0),
\eea
where the rank of $i:\, H^1(X, Q_1) \rightarrow H^2(X, V_a \otimes V_b)$ can be in the range $[0,2]$. Thus we have at least $H^*(X,{\textstyle \bigwedge}^2 V_1) = (0,51,51,0)$.

In the hidden sector, the cohomology is
\bea
        H^*(X,V_2) &=& (0,0,0,0),\nonumber\\
        H^*(X,V_2^\vee\otimes L^{-1}) &=& (0, 2-\rank j, 2-\rank j, 0),
\eea
where the rank of $j:H^1(\sigma, {\cal N}_{V_2}\otimes K_B \otimes {\cal O}(-C)|_\sigma) \rightarrow H^1 (\sigma, {\cal N}_{V_2}\otimes K_B |_\sigma)
$ can again lie at most within the range $[0,2]$.

A remark about the actual ranks of the linear maps $f,g,h,i,j$ is in order. Their concrete value depends on the choice of bundle moduli and can therefore vary over the moduli space. To decide which values they can really take within the naive ranges stated above requires a more thorough analysis as performed, in the context of $SU(N)$ bundles, in \cite{Donagi:2004ia}. Since it is of phenomenological relevance, we restrict our attention here to the rank of the map $j$, which decides about the appearance of possible exotic matter in the form of extra right-handed electrons.
A detailed, but straightforward analysis along the lines of \cite{Donagi:2004ia} reveals that the possible values for $\rank j$ are $0$ and $2$, with $2$ being the generic value and $0$ corresponding to a specific choice of bundle moduli for $V_2$. We therefore restrict ourselves to the generic maximal value leading indeed to $H^*(X,V_2^\vee\otimes L^{-1}) = (0,0,0,0)$, as desired. For this generic choice of moduli the number of Higgses is then in the range $[51,55]$ and can be determined in a similar manner, though we do not perform this analysis here.

To conclude, we list again the total spectrum of our example in table \ref{spec_gut_ex}.
\begin{table}[htb]
\renewcommand{\arraystretch}{1.5}
\begin{center}
\begin{tabular}{|c||c|c|c|}
\hline
\hline
$SU(5)\times U(1)_X \times E_6$ & Cohomology & $\chi$  & SM part. \\
\hline \hline
$ ({\bf 10},{\bf 1})_{1\over 2}$ & $(0,1,4,0)$ & $3$ & $(q_L,d^c_R,\nu^c_R)+[H_{10}+\overline H_{10}]$ \\
$( {\bf 10},{\bf 1})_{-2}$ & $(0,0,0,0)$ & $0$ &$-$ \\
\hline
$(\overline \5,{\bf 1})_{-{3\over 2}}$ & $(0,0,3,0)$ & $3$ & $(u^c_R,l_L)$ \\
$(\overline \5,{\bf 1})_{1}$ & $(0,[51,55],[51,55],0)$ & $0$ & $[(h_3, h_2)+(\overline h_3, \overline h_2)]$ \\
\hline

 $({\bf 1},{\bf 1})_{{5\over 2}}$ & $(0,0,3,0)$ & $3$ & $e^c_R$ \\
\hline
$({\bf 1},{\bf 27})_{{5\over 6}}$ & $(0,0,0,0)$ & $0$ & $-$ \\
$({\bf 1},{\bf 27})_{-{5\over 3}}$ & $(0,0,0,0)$ & $0$ & $-$ \\
\hline
\end{tabular}
\caption{\small Massless spectrum of a flipped $SU(5)$ model with hidden $E_6$ symmetry. 
     }
\label{spec_gut_ex}
\end{center}
\end{table}

\section{Conclusions}

In this paper we have provided the technical tools for the 
computation of the complete massless spectrum of 
heterotic string compactifications invoking vector bundles
with $U(N)$ structure groups. Our main results are both of purely mathematical interest and lead, from the physical point of view, to the construction of new quasi-realistic heterotic string compactifications.

Taking the one-loop corrections to the Donaldson-Uhlenbeck-Yau
equation derived in \cite{Blumenhagen:2005ga} seriously, we have 
proposed a new notion of stability for the loop and
non-perturbatively corrected Hermitian Yang-Mills equation.
It is the analogue of the concept of $\Pi$-stability of B-type
D-branes \cite{Douglas:2000ah} for vector bundles in the $E_8\times E_8$
heterotic string. While, in the context of importance to us in this publication, this modified stability concept reduces to the familiar one of $\mu$-stability, it would be interesting to investigate the implications of $\Lambda$-stability both from the mathematical point of view and with respect to applications in string model building.  

In the technical main part of this article we have extended the results of \cite{Donagi:2004ia} 
concerning the computation
of cohomology groups for vector bundles defined via the spectral cover
method. In particular, we have provided the expressions for
$H^i(X,V_a\otimes V_b)$, $H^i(X, \bigwedge^2 V)$  and
$H^i(X, {\bf S}^2 V)$, where for the latter two
our results differ significantly from the ones obtained
in \cite{Donagi:2004ia}. In all these cases the cohomology can be computed
from certain line bundles living on the intersection curves
of the two spectral cover surfaces involved. Therefore,
eventually the technical computation of the massless spectrum
boils down to the determination of the cohomologies
of line bundles on certain curves.
For $U(4)$ bundles defined via non-split extensions of two $U(2)$ bundles,
we have provided an explicit  proof for their $\mu$-stability. 
 
In the remaining, more physically oriented part of this paper we have applied all these techniques to the construction of
stringy flipped $SU(5)$ models as proposed in \cite{Blumenhagen:2006ux},
where the masslessness of the $U(1)_X$ introduced
additional constraints on the $SU(4)\times U(1)$ bundle
involved. Defining the $U(4)$ bundle via an extension
of two $U(2)$ bundles, we have found what we believe is the first  fully consistent, supersymmetric flipped $SU(5)$ string model with 
just the MSSM matter spectrum, i.e. without any additional
vector-like matter. Moreover, this model exhibits precisely one vector-like pair of
the desired GUT Higgs fields in the antisymmetric
representation of $SU(5)$ allowing for field theoretic GUT symmetry breaking down to the Standard Model gauge group. The only major shortcoming
is the appearance of a large number of electroweak Higgs fields. The common philosophy how to deal with unwanted vector-like pairs would be to carefully analyse their mass matrix and determine whether they can acquire sufficiently large masses as to decouple from the effective low-energy theory (for recent examples in the heterotic literature see e.g. \cite{Buchmuller:2006ik,Faraggi:2006qa,Lebedev:2006kn,Lebedev:2006tr}). We leave such an analysis for future work, but hasten to stress that the extra Higgs pairs are a  consequence of the very specific geometric background and the types of vector bundles employed and may be avoidable in different setups. 

As discussed in \cite{Blumenhagen:2006ux}, for the type of flipped SU(5) vacua studied in this article there are no obvious selection rules forbidding any of the observed Yukawa couplings, whereas potentially problematic dimension four, five and six operators inducing unacceptable proton decay are absent. We consider this latter issue as a clear phenomenological advantage which is known to distinguish flipped SU(5) from the Georgi-Glashow GUT scenario. Further phenomenological studies would involve the actual computation of the relevant interaction terms including the ones involving the GUT Higgs and which are required for the field theoretic symmetry breaking. 

As an alternative to this type of GUT model building, it would be interesting to apply the methods of this paper to the construction of string vacua directly with MSSM gauge group, along the lines of \cite{Blumenhagen:2006ux}. These vacua are defined by embedding a vector bundle of structure group $SU(5) \times U(1)$ into $E_8$, yielding $SU(3) \times SU(2) \times U(1)_Y$ in four dimensions. We plan to come back to these questions in the future.

What we find most interesting in the light of recent discussions concerning the gauge sector of the four-dimensional string landscape (e.g. \cite{Blumenhagen:2004xx,Dijkstra:2004cc,Gmeiner:2005vz,Dienes:2006ut,Douglas:2006xy,Anastasopoulos:2006da,Lebedev:2006tr}) is the fact that the model presented in this article is just one example of a much larger class of heterotic vacua which, as we recall, are defined on general simply-connected Calabi-Yau manifolds. In particular, they do not rely on highly non-trivial properties of the fundamental group of the internal space or on full solvability of the underlying CFT.  
We are quite confident 
that by extending the analysis of this paper to more generic backgrounds and vector bundles, models with just the MSSM spectrum can be found.

\vskip 1cm
 {\noindent  {\Large \bf Acknowledgements}}
 \vskip 0.5cm 
We gratefully acknowledge  helpful discussions with  Volker Braun, Emanuel Diaconescu, Ron Donagi, Florian Gmeiner, Yang-Hui He, Gabriele Honecker,
Peter Mayr and useful correspondence with Alexander Degtyarev.  
R.B. would like to thank the physics department of Rutgers University
for hospitality. The work of R.R. was supported in part by DOE grant DE-FG02-96ER40959.
The work of T.W. was supported in part by
DOE grant
DOE-EY-76-02-3071.
 \vskip 2cm

\appendix

\section{Proof of stability for rank four bundles $V$ by two generic rank two bundles $V_1$, $V_2$}
\label{proof}

Let $X$ be a generic elliptically fibered Calabi-Yau manifold as considered in section \ref{seccy}. In particular we assume that $X$ admits exactly one section \footnote{Otherwise, we assume that $\wedge^{2}V_{2}$ restricted to a generic fiber is trivial.}.
Consider a vector bundle $V$ defined by the short exact sequence
\begin{equation}
\label{exten}
0 \to V_1 \to V \to V_2 \to 0,
\end{equation}
where $V_i$ are $U(N)$ or $SU(N)$ bundles over $X$ that correspond to two different generic irreducible spectral covers. This implies that the restriction of $V_i$ to a generic fiber is isomorphic to the sum of degree zero line bundles which  are mutually different and that $V_i$ is stable with respect to  any ample class of the form $J=J_{X}+n\pi^* J_B$ for sufficiently large $n$. Here  $J_{X}$ and $J_{B}$  denote  ample classes on $X$ and $B$,  respectively.

In the sequel, we restrict ourselves to the case that  both $V_i$ are of rank $2$, as in section \ref{sec_GUT}. 
Consider a bundle $V$ that corresponds to a non-trivial element in $Ext^1(V_2,V_1)$ and satisfies $\mu(V_1)< \mu(V)<\mu(V_2)$ with $\mu(V) =0$. We will show that under these assumptions $V$ is a stable bundle with respect to any ample class of the form $J=J_{X} +n\pi^* J_B$ for sufficiently large $n$. In particular, we have to show that all torsion free sheaves ${\cal N}$ of rank smaller than four which admit an injective map 
$$
{\cal N} \to V 
$$
obey $\mu({\cal N})<0$. Note that, as discussed for example in \cite{Donagi:2006yf}, it is sufficient to show this statement for all vector bundles of rank smaller than four.

Since the restriction of $V$ to a generic fiber is by construction  isomorphic to  the sum of mutually different degree zero line bundles, the degree of all subbundles of $V$ along the generic fiber is smaller than or equal to zero.
As can be seen by straightforward computation, for subbundles of degree smaller than zero along the generic fiber, large $n$ is sufficient to make their slope negative, hence they cannot destabilize $V$  \cite{Friedman:1997yq}. 

Therefore we have to consider only subbundles of degree zero along the generic fiber. The fact that the spectral cover of $V$ is the union of the two irreducible spectral covers of $V_{1}$ and $V_{{2}}$ implies that these subbundles are of rank two.

Instead of checking for destabilising rank two subbundles of $V$, we can check for destabilising sub line bundles of ${\textstyle \bigwedge}^2 V$ \footnote{This is a special case of the general fact that for a rank $r $ subbundle $W_r$ of a rank $m$ bundle $V_m$ also ${\textstyle \bigwedge}^r W_r \subset {\textstyle \bigwedge}^r V_m$.}.  For this purpose we make use of the fact that ${\textstyle \bigwedge}^2 V$ fits into the exact sequence
\begin{equation}
\xymatrix{
                              &                             & {\textstyle \bigwedge}^2 V_2\\
{\textstyle \bigwedge}^2 V_1 \ar[r]& {\textstyle \bigwedge}^2 V \ar[r] &Q \ar[u]\\
                              &                             & V_1 \otimes V_2\ar[u]
}
\end{equation}
It follows that  the subbundles ${\cal N}$ of ${\textstyle \bigwedge}^2 V$ are either subbundles of ${\textstyle \bigwedge}^2 V_1$ or subbundles of $Q$ lifting to ${\textstyle \bigwedge}^2 V$.
In the first case these bundles cannot be destabilising  since 
$$
 \mu({\cal N}) \leq \mu({\textstyle \bigwedge}^2 V_1) < \mu( {\textstyle \bigwedge}^2 V).
$$ 
Note that the first inequality is due to fact that the rank of ${\textstyle \bigwedge}^2 V_1$ is one and hence the cokernel of ${\cal N} \rightarrow {\textstyle \bigwedge}^2 V_1$ is a torsion sheaf.
It remains to show that subbundles of $Q$ with non-negative slope do not lift to ${\textstyle \bigwedge}^2 V$.  

Every line bundle which is a subbundle of $Q$ must either be a subbundle of $V_1 \otimes V_2$ or a subbundle of ${\textstyle \bigwedge}^2 V_2$ lifting to $Q$.
However,  for generic spectral covers of $V_1$ and $V_2$, $V_1 \otimes V_2$ itself corresponds to an irreducible spectral cover and therefore has no subbundles of rank one and degree zero along the fiber, as discussed above.

Turning to the second possibility, we
note that it follows from our assumptions that ${\textstyle \bigwedge}^2 V_2 = \pi^* L_2$ for some line bundle $L_2$ on $B$. 
Consider subbundles of $\pi^* L_2$ of degree zero along the fiber. They are of the form $\pi^*D$ for some line bundle $D$ on $B$. In order for them to  destabilize ${\textstyle \bigwedge}^2 V$ they have to lift to ${\textstyle \bigwedge}^2 V$, hence they have to lift to $Q$. 

We will show that this is impossible. 
As a standard matter of fact, every diagram 
\begin{equation}
\xymatrix{
V_1\otimes V_2 \ar[r] & Q \ar[r]                       & \pi^* L_2 \\
                              &                             & \pi^*D \ar[u]
}
\end{equation}
can be completed to 
\begin{equation}
\label{complete}
\xymatrix{
                              &                             & \pi^*F\\
V_1\otimes V_2 \ar[r]        & Q \ar[r]                 & \pi^* L_2 \ar[u] \\
V_1 \otimes V_2 \ar[r] \ar[u]     & Q' \ar[r] \ar[u]              & \pi^*D \ar[u]
}
\end{equation}
for some sheaf $F$ with support on a divisor $S$ on $B$ (see e.g. Chapter III of \cite{MacLane}).
It is easy to see that in our specific case $F$ is a line bundle on $S$.
In addition $\pi^*D$ lifts to $Q$ if and only if $Q'$ corresponds to the trivial extension, i.e. $Q^{'}=0 \in Ext^1(\pi^*D,V_1\otimes V_2)$ \cite{MacLane}.
We can assume that $Q$ is not the trivial extension.
There exists a natural map 
$$
Ext^1(\pi^*L_2,V_1\otimes V_2) \rightarrow Ext^1(\pi^*D,V_1\otimes V_2)
$$
and we can complete our proof by showing  that this map is an injection.

To do so consider the short exact sequence
\begin{equation}
0 \to \pi^* D \to  \pi^* L_2 \to \pi^* F \to 0
\end{equation}
in (\ref{complete}) inducing the long exact sequence
\begin{equation}
\cdots \to Ext^1(\pi^*F,V_1\otimes V_2) \to Ext^1(\pi^*L_2,V_1\otimes V_2) \to Ext^1(\pi^*D,V_1\otimes V_2) \to \cdots
\end{equation}
We conclude that a sufficient condition for $Q'$ not being the trivial extension is the vanishing of $Ext^1(\pi^*F,V_1\otimes V_2)$. 
Consider
\begin{eqnarray}
Ext^1_{X}(\pi^*F,V_1\otimes V_2)&=&Ext_{X}^2(V_1\otimes V_2,\pi^*F)^{*}\\
&=& H^{2}(X,\pi^*F\otimes V_1^{*}\otimes V_2^{*})^{*}\\
&=& H^2(\pi^*S,V_1^{*}\otimes V_2^{*}\otimes \pi^*F)\\
&=&H^0(\pi^*S,V_1\otimes V_2 \otimes \pi^*F^{*} \otimes K_{\pi^* S}) \\ 
&=& H^0(S,\pi_*(V_1\otimes V_2 \otimes K_{\pi^* S}) \otimes F^{*})=0,
\end{eqnarray}
where we use Serre duality on $X$ and on $\pi^*S$. The last equality follows from the fact that $\pi_*(V_1 \otimes V_2)=0$ for generic spectral cover bundles $V_i$.

\section{Cohomology of line bundles over del-Pezzo surfaces}
\label{app_dPr}
In order to determine the  cohomology classes of line bundles
over general del Pezzo surfaces $dP_r$, $r=0,\ldots,8$ we proceed as
follows. We will first compute the effect of blowing up just a single
point on $\IP_2$ and will then argue  that the different blow-ups
are independent of each other. This leads directly to the general formula
for $r$ blown up points.

Blowing up just a single point results in $dP_1$, which is the same
as the Hirzebruch surface $F_1$.  The latter is a $\IP_1$ fibration
over $\IP_1$ and we can therefore apply the Leray spectral sequence
for this fibration $dP_1=F_1 \stackrel{\pi}{\to} \IP_1$.
More concretely, consider a line bundle on $dP_1$ with  first Chern
class
\bea
          c_1(L)=a\, l + b\, E_1 = a\, (l-E_1) +(b+a)\, E_1,
\eea
where $S=E_1$ and ${\cal E}=l-E_1$ are precisely the two $\IP_1$s
appearing in $F_1$. The intersection form for these 2-cycles is
\bea
        S\cdot S=-1, \quad S\cdot {\cal E}=1, \quad {\cal E}\cdot {\cal E}=0.
\eea
As has been shown in \cite{Donagi:2004ia} and can be verified by utilizing
for instance the Grothendieck-Riemann-Roch theorem the push-forward
of a line bundle onto the $\IP_1$ described by the divisor $S$
is
\bea
    \pi_*(L)&=&{\cal O}(a{\cal E})\otimes\left[ {\cal O}\oplus
             {\cal O}(-{\cal E})\oplus\ldots\oplus 
           {\cal O}(-(a+b){\cal E})\right]\, \quad {\rm for}\ a+b\ge 0\\
    R^1\, \pi_*(L)&=&{\cal O}(a{\cal E})\otimes\left[ 
             {\cal O}({\cal E})\oplus {\cal O}(2{\cal E})\oplus\ldots\oplus 
           {\cal O}(-(a+b+1){\cal E})\right]\, \quad {\rm for}\ a+b< 0. \nonumber
\eea
Applying now Bott's formula for the cohomology classes of line bundles
on $\IP_1$ gives the cohomology classes of the push-forward line bundles
on $\IP_1$,
\bea
         H^0(\IP_1,\pi_* L)=\begin{cases}
                          { a+2 \choose 2} & {\rm for}\ a\ge 0,\ b\ge 0 \\
                          { a+2 \choose 2}-{ b \choose 2} & 
                                 {\rm for}\  a\ge 0,\ -a\le b<0 \\
                            0 & {\rm else} \\ 
                             \end{cases}
\eea
and
\bea
         H^1(\IP_1,\pi_* L)=\begin{cases}
                          { b \choose 2} & {\rm for}\ a\ge 0,\ b\ge 0 \\
                      -{ a+2 \choose 2}+{ b \choose 2} & {\rm for}\  
                  a<0,\ b>-a \\
                        0 & {\rm else}. \\ 
                             \end{cases}
\eea
Similarly, for the cohomology classes of the first right derived functor
we find
\bea
         H^0(\IP_1,R^1\, \pi_* L)=\begin{cases}
                          -{ a+2 \choose 2}+{ b \choose 2} & 
                   {\rm for}\ a\ge 0,\ b< -a \\
                          { b \choose 2} & 
                                 {\rm for}\  a<0,\ b<0 \\
                            0 & {\rm else} \\ 
                             \end{cases}
\eea
and
\bea
         H^1(\IP_1,R^1\, \pi_* L)=\begin{cases}
                          { a+2 \choose 2} & 
                   {\rm for}\ a<0,\ b<0 \\
                     { a+2 \choose 2}- { b \choose 2}& 
                   {\rm for}\ a<0,\ 0<b<-a \\
                          0 & {\rm else}. \\
                             \end{cases}
\eea
With the help of the Leray spectral sequence it is now straightforward to compute 
$H^i(dP_1,L)$. Using the above decoupling argument for the different
blow-ups the final result for general del-Pezzo surfaces $dP_r$
can be written in the following suggestive form.
Consider the general line bundle on $dP_r$ with
\bea
   c_1(L)=a_0\, l +\sum_{i=1}^\rho b_i\, E_i + \sum_{j=\rho+1}^r c_j\, E_j
   \ {\rm with}\ b_i<0\ {\rm and}\  c_j\ge 0. 
\eea
For $a_0\ge 0$ define
\bea
        A={ a_0+2\choose 2} - \sum_{i=1}^\rho {b_i \choose 2}. 
\eea
If $A\ge 0$ the cohomology classes of the line bundle are
\bea
     H^*(dP_r,L)=\left( A, \sum_{j=\rho+1}^r { c_j \choose 2} , 0\right)
\eea
and for $A<0$ they are
\bea
     H^*(dP_r,L)=\left(0, \sum_{j=\rho+1}^r { c_j \choose 2} -A, 0\right).
\eea
Similarly, if $a_0<0$ we define
\bea
        A={ a_0+2\choose 2} - \sum_{i=1}^\rho {c_j \choose 2}. 
\eea
If $A\ge 0$ the cohomology classes of the line bundle are
\bea
     H^*(dP_r,L)=\left( 0, \sum_{i=1}^\rho { b_i \choose 2} , A\right)
\eea
and for $A<0$ they are
\bea
     H^*(dP_r,L)=\left(0, \sum_{i=1}^\rho { b_i \choose 2} -A, 0\right).
\eea
Of course these formulae are consistent with the Riemann-Roch-Hirzebruch
formula for the Euler characteristic of these line bundles over $dP_r$.
In addition we have checked that for the toric del-Pezzo surfaces
$dP_0, \ldots, dP_3$ they are consistent with the cohomology classes
derived using toric methods.

\section{Koszul sequence for $H^*(X, V_a \otimes V_b)$ }
\label{appKos}

As derived in section (\ref{sec_VaVb}), the cohomology groups of the tensor product of two spectral cover bundles $V_a$ and  $V_a$ are given by
\bea
H^i(X, V_a \otimes V_b) = H^{i-1}(C_a \cap C_b, {\cal L}|_{C_a \cap C_b})\,\, \,   {\rm for}\,  i=1,2,
\eea
where ${\cal L} = {\cal N}_{V_a} \otimes {\cal N}_{V_b} \otimes K_B$. Our task is thus to compute the cohomology $H^*(C_a \cap C_b, {\cal L}|_{C_a \cap C_b})$  for a  line bundle  ${\cal L}$  defined on the elliptically fibered three-fold $X$. This can be accomplished by invoking the  Koszul sequence

\bea
\label{Koszul1}
(I) \quad 0 \rightarrow \cL \otimes {\cal O}(- C_a)|_{C_b} \rightarrow \cL|_{C_b} \rightarrow \cL|_{C_a \cap C_b}
\rightarrow 0 \ .
\eea
The point is that each of the first two objects can again be computed from known objects on $X$ via a Koszul sequence of its own, 

\bea
\label{Koszul2}
(II) \quad 0 \rightarrow \cL \otimes {\cal O}(- C_b) \rightarrow \cL  \rightarrow \cL|_{C_b}
\rightarrow 0 \ 
\eea
and
\bea
\label{Koszul3}
(III) \quad 0 \rightarrow \cL \otimes {\cal O}(- C_a -C_b) \rightarrow \cL \otimes {\cal O}(-C_a)  \rightarrow \cL \otimes {\cal O}(-C_a)|_{C_b}
\rightarrow 0 \ .
\eea

\noindent Each of these three  short exact sequences induces a long exact sequence in cohomology. 

We therefore need as our input data the dimensions of the cohomology groups of the four line bundles on $X$
\bea
H^*(X,\cL) &=&  H^*(X,{\cal N}_{V_a} \otimes {\cal N}_{V_b} \otimes K_B), \nonumber \\
H^*(X,\cL \otimes {\cal O}(-C_a)) &=&  H^*(X,{\cal N}_{V_a} \otimes {\cal N}_{V_b} \otimes K_B \otimes {\cal O}(-C_a)),  \\
H^*(X,\cL \otimes {\cal O}(-C_b)) &=&  H^*(X,{\cal N}_{V_a} \otimes {\cal N}_{V_b} \otimes K_B \otimes {\cal O}(-C_b)), \nonumber \\
H^*(X,\cL \otimes {\cal O}(-C_a-C_b)) &=&  H^*(X,{\cal N}_{V_a} \otimes {\cal N}_{V_b} \otimes K_B \otimes {\cal O}(-C_a-C_b)). \nonumber
\eea

These can easily be obtained with the help of the general expressions for the cohomology groups of line bundles on $X$ given in section 3.1 together with appendix B.

\clearpage
\nocite{*}
\bibliography{rev}
\bibliographystyle{utphys}

\end{document}